\documentclass[a4,11pt]{article}

\usepackage{amsmath,amssymb,color,graphics,graphicx,amscd,amsfonts,epsf,ulem}
\def\L{\Lambda}
\def\th{\theta}
\def\mN{\mathcal{N}}
\def\m{\mu}
\def\n{\nu}
\def\Tr{{\rm Tr}}
\def\th{\theta}
\def\e{\epsilon}
\def\a{\alpha}
\def\b{\beta}
\def\c{\gamma}

\def\e{\epsilon}
\def\r{\rho}
\def\C{\Gamma}
\def\der{\partial}
\def\hk{\hat{k}}
\def\tk{\tilde{k}}

\allowdisplaybreaks[4]

\date{}
\begin{document}

\begin{flushright} 
%\today\\
SU-ITP-14-25\\ 
OIQP-14-8\\
\end{flushright} 

\vspace{0.1cm}

\begin{center}
  {\LARGE
  
  On the continuity of the commutative limit
   of the 
   
4d ${\cal N}=4$ non-commutative super Yang-Mills theory
   
  }
\end{center}
\vspace{0.1cm}
\vspace{0.1cm}
\begin{center}

         Masanori H{\sc anada}$^{a,b,c}$ 
and  
Hidehiko S{\sc himada}$^{d}$

\vspace{0.3cm}

\normalsize{$^{a}$Stanford Institute for Theoretical Physics,
Stanford University, Stanford, CA 94305, USA}\\
\normalsize{$^{b}$Yukawa Institute for Theoretical Physics, Kyoto University,}\\
\normalsize{Kitashirakawa Oiwakecho, Sakyo-ku, Kyoto 606-8502, Japan}\\
\normalsize{$^{c}$The Hakubi Center for Advanced Research, Kyoto University,}\\
\normalsize{Yoshida Ushinomiyacho, Sakyo-ku, Kyoto 606-8501, Japan}\\

${}^{d}$
{\normalsize Okayama Institute for Quantum Physics, Okayama, Japan}\\

\end{center}

%\newpage

\vspace{1.5cm}

\begin{center}
  {\bf abstract}
\end{center}

We study the commutative limit of the non-commutative 
maximally supersymmetric Yang-Mills theory in four dimensions ($\mathcal{N}=4$ SYM). 
The commutative limits of non-commutative spaces are important
in particular in the applications of non-commutative spaces for regularisation
of supersymmetric theories (such as the use of non-commutative spaces
as alternatives to lattices for supersymmetric gauge theories and 
interpretations of some matrix models as regularised 
supermembrane or superstring theories), which 
in turn 
can play  
a prominent role 
in the study of quantum gravity via the gauge/gravity duality. 
In general, the commutative limits 
are known to be singular and non-smooth  
due to UV/IR mixing effects. 
We give a direct proof that UV effects 
do not break the
continuity of the commutative limit of the 
non-commutative $\mathcal{N}=4$ SYM to all order in perturbation theory,
including non-planar contributions.  
This is achieved by 
establishing the uniform convergence (with respect to the 
non-commutative parameter) of momentum integrals associated with 
all Feynman diagrams appearing in the theory,
using the same tools involved in the 
proof of finiteness of the commutative $\mathcal{N}=4$ SYM.

%%%%%%%%%%%%%%%%%%%%%%%%%%%%%%%%%%%%%%%%%%%%%%%%%%%%%%%%%%%%%%%%%%%%%%
%%%%%%%%%%%%%%%%%%%%%%%%%%%%%%%%%%%%%%%%%%%%%%%%%%%%%%%%%%%%%%%%%%%%%%
%%%%%%%%%%%%%%%%%%%%%%%%%%%%%%%%%%%%%%%%%%%%%%%%%%%%%%%%%%%%%%%%%%%%%%
\newpage

%%%%%%%%%%%%%%%%%%%%%%%%%%%%%%%%%%%%%%%%%%%%%%%%%%%%%%%%%%%%%%%%%%%%%%
%%%%%%%%%%%%%%%%%%%%%%%%%%%%%%%%%%%%%%%%%%%%%%%%%%%%%%%%%%%%%%%%%%%%%%
%%%%%%%%%%%%%%%%%%%%%%%%%%%%%%%%%%%%%%%%%%%%%%%%%%%%%%%%%%%%%%%%%%%%%%
\section{Introduction}
\hspace{0.51cm}
%%%%%%%%%%%%%%%%%%%%%%%%%%%%%%%%%%%%%%%%%%%%%%%%%%%%%%%%%%%%%%%%%%%%%%
%%%%%%%%%%%%%%%%%%%%%%%%%%%%%%%%%%%%%%%%%%%%%%%%%%%%%%%%%%%%%%%%%%%%%%
%%%%%%%%%%%%%%%%%%%%%%%%%%%%%%%%%%%%%%%%%%%%%%%%%%%%%%%%%%%%%%%%%%%%%%
Non-commutative field theories are non-local deformations 
of usual local field theories, obtained by replacing products between 
fields by the so-called Moyal products, 
\begin{eqnarray}
f \star g
\equiv
f
e^{\frac{i}{2} \overset{\leftarrow}{\partial}_\mu C^{\mu\nu}\overset{\rightarrow}{\partial}_\nu}
g   
=f g 
+ 
\frac{i}{2}
C^{\mu\nu} 
(\partial_\mu f) 
(\partial_\nu g)
+
\frac{1}{2!}\left(\frac{i}{2}\right)^2 
C^{\mu_1 \nu_1}
C^{\mu_2 \nu_2}
( \partial_{\mu_1} \partial_{\mu_2} f)
 (\partial_{\nu_1} \partial_{\nu_2} g)
+ \cdots,
\label{RFMoyal}
\end{eqnarray} 
where $C^{\mu\nu}=-C^{\nu\mu}$ are the non-commutativity parameters.  
Aspects of these theories have been studied from various perspectives
in recent years. For a review, see e.g. \cite{Douglas:2001ba}. 

In this paper we study the commutative limit, $C \rightarrow 0$.  
The original local field theories are recovered in this limit at the classical level.
However,
the commutative limit is known to be singular
at the quantum level for generic non-commutative field theories,
due to an effect called the ``UV/IR mixing'' 
\cite{Minwalla:1999px}, as will be elaborated later.

One motivation to study the non-commutative field theory,
or the non-commutative space, 
comes from the expectation that it might provide us with a good mean to regularise, or discretise,  
quantum theories with infinite degrees of freedom (in particular those with supersymmetry), 
enabling us to define these theories non-perturbatively. 
The commutative limit plays a crucial role in this context~\footnote{
The commutative limit is also important, if one 
pursues the possibility that our space-time is non-commutative. 
This non-commutativity is not observed so far.
If the commutative limit is continuous, this can be naturally attributed to  
the smallness of the value of $C$. 
If the commutative limit is singular,
which is the case for generic non-commutative field theories,
it is more difficult to explain the absence of the non-commutativity
in the present observation. 
}.

One of the early examples 
of the application of the non-commutative space
is the construction of the matrix model of M-theory 
as the regularised version of supermembrane 
theory\cite{RBGoldstone,RBHoppeThesis,RBdWHN,RBBFSS}. 
(The matrix model has the same amount of supersymmetry as the supermembrane theory; 
an important advantage of the regularisations using the non-commutative spaces compared to, 
for example,
simple lattice regularisations is that the 
supersymmetry can often be preserved more easily.)
The mathematical structures associated with this regularisation are 
the same as those appearing in the non-commutative spaces.
It is an important issue to understand how one should take the large-$N$ limit
of the matrix model at the quantum level, which can be interpreted as  
the continuum limit of the membrane theory.
This continuum limit of the membrane theory is equivalent to 
the commutative limit $C\to 0$,  
in the special case where the membrane worldspace 
(the timeslice of the membrane worldvolume) is given by 
the so-called non-commutative plane defined 
by the Moyal product (\ref{RFMoyal}).~\footnote{
Usually, compact worldspaces of membranes (such as a sphere),
whose matrix counterparts are finite dimensional matrices,
are considered.
The matrix version of a non-compact worldspace by contrast is 
infinite dimensional because of infinite numbers
of degrees of freedom in the IR. 
Strictly speaking, for the case of infinite dimensional matrices corresponding to
the non-compact membranes,
the quantum theory is potentially ill-defined 
due to the infinite number of degrees of freedom, 
and hence one has to
consider it as a certain limit of the theory associated with finite-dimensional matrices.
Nonetheless, we believe that 
at least some of the essential features of the continuum limit of membranes
should be captured by the $C \to 0$ limit of the non-commutative plane.
}  
The IKKT matrix model~\cite{RBIKKT} is also  
obtained by applying a similar regularisation to the superstring worldsheets.

Another example, which is more directly relevant to the 
subject of this paper, 
is the application of non-commutative spaces 
to non-perturbative definitions of supersymmetric Yang-Mills (SYM) theories. 
The non-perturbative definition of SYM theories 
via regularisations of them is of course a conceptually important theme,
and also opens up the possibility 
of studying non-perturbative properties of 
these theories via Monte-Carlo simulations. 
However, construction of satisfactory formulations 
of regularised SYM theories (in particular those using the lattices) 
is a notoriously difficult problem,
whose general solution is not known to date.
In general, one cannot preserve the full supersymmetry algebra 
in the regularised theory. It is possible to write down a discretised
theory which recovers the supersymmetry in the continuum limit at tree level;
however, if one goes beyond the tree level, 
it is in general necessary to 
introduce counter-terms to prevent the explicit breaking of the 
supersymmetry via radiative corrections.
This procedure is usually called as the fine-tuning.
Only for some specific SYM theories, 
lattice regularisation 
methods which avoid the fine-tuning problem are known.

A particularly important four-dimensional SYM theory 
is that with the maximal amount of supersymmetry, 
${\cal N}=4$ supersymmetric Yang-Mills theory
(4d ${\cal N}=4$ SYM hereafter), 
whose non-perturbative properties are studied extensively 
in particular in the context of
the AdS/CFT correspondence \cite{RBMaldacena}. 
For the $\mathcal{N}=4$ SYM, no formulation based purely on the lattice 
regularisation is available which is free from the 
fine-tuning problem\footnote{
Lattice simulation 
of $\mathcal{N}=4$ SYM, albeit with the parameter fine tuning, 
is 
also pursued. See \cite{Catterall:2014vka} for the latest result. 
}. 
Several years ago, one of the authors proposed a fine-tuning 
free non-perturbative regularisation of 
the theory utilising the non-commutative space \cite{Hanada:2010kt}\footnote{
Prior to this work, a similar technique has been applied for 
3d maximal SYM in \cite{Maldacena:2002rb}. 
See also \cite{Unsal:2005us}, in which all dimensions are embedded into matrices. 
Another example which uses non-commutative space can be found in \cite{Ishii:2008ib}.  
}. 
In this formalism, two spatial dimensions are 
non-commutative, being embedded into the gauge degrees of freedom 
\cite{Aoki:1999vr,Myers:1999ps}. 
The remaining two directions are regularised by the lattice method\footnote{
For two-dimensional super Yang-Mills, 
there are several proposals of fine-tuning free 
formulations \cite{Kaplan:2005ta, Hanada:2010kt}, and   
there are numerical tests which support the validity of 
these proposals at 
the
nonperturbative level \cite{Kanamori:2008bk}. 
For a complete list of references, see a review paper \cite{Catterall:2009it}. 
}. 
According to this proposal, after an appropriate continuum limit is taken, 
one obtains the non-commutative version of the 4d $\mathcal{N}=4$ SYM
(in which only two spacelike directions are non-commutative). 
A crucial assumption here is that 
4d ${\cal N}=4$ SYM is obtained as the commutative limit of its non-commutative cousin.  

However, 
it is a non-trivial issue whether
one recovers the original theory by taking the commutative limit.
An important feature of generic non-commutative field theories 
is that computations of some Feynman diagrams, 
whose counterparts in the commutative theory are UV-divergent, 
yield terms which behave singularly in the $C\to 0$ limit.
(These terms behave singularly  
also in the limit where the external momenta go to zero,
and hence the appearance of these terms is usually called as the
``UV/IR mixing'' \cite{Minwalla:1999px}.)
This implies 
that in the commutative limit the observables of 
a non-commutative field theory 
are not equivalent to those of its commutative counterpart, 
at least without further modification of the theory.  
In this paper we will show that  
the breaking of the continuity of the commutative limit 
due to UV effects does not occur 
for the four-dimensional $\mathcal{N}=4$ SYM.  
\footnote{
In this paper, we avoid the introduction of
non-commutativity between the timelike and a spacelike coordinate.
It has been argued that introduction of the non-commutativity between the time direction 
and a spacelike direction leads to pathological features, 
such as the violation of the causality and unitarity\cite{Seiberg:2000gc,AlvarezGaume:2000bv,Gomis:2000zz}.
Furthermore,
for the regularisation of $\mathcal{N}=4$ SYM 
in the approach of \cite{Hanada:2010kt}, the non-commutativity is introduced   
only between two spacelike dimensions. 
It is 
the commutative limit in this setting
that is crucial 
in this approach.}
More precisely, we prove that
the commutative limit is continuous, for all Green functions
of the non-commutative ${\cal N}=4$ SYM in the lightcone gauge,
to all order in the perturbation theory, including non-planar contributions.
Actually, it was suggested in \cite{Matusis:2000jf}  that the 
singular terms
are absent for the non-commutative ${\cal N}=4$ SYM, which implies
the continuity of the $C\rightarrow 0$ limit. 
One reason behind this suggestion is the
well-known finiteness of the commutative ($C=0$)
$\mathcal{N}=4$ SYM\cite{RBMandelstam,RBBrinkLindgrenNilssonProof}. 
However, the finiteness of the commutative theory alone 
does not ensure the continuity, let alone the smoothness,
of the commutative limit, as will be explained in detail 
in section \ref{RSNontrivial}.
The crucial point is that the finiteness of the original
theory is due to non-trivial cancellations between divergent
diagrams, and there is no guarantee a priori 
that these cancellations are not ruined by the 
introduction of the  
non-commutativity.

The crucial concept utilised in our proof is the uniform convergence,
whose relevance is explained in section \ref{RSNontrivial}.
Our tools 
to prove the uniform convergence
of all Feynman integrals,
the lightcone superspace and power-counting procedures done in two steps, 
are those used in
the original proofs of the finiteness of the commutative 
theory~\cite{RBMandelstam,RBBrinkLindgrenNilssonProof}.
We will prove our theorem by showing
that these tools remain effective after modifications due to the
non-commutativity.
This strategy is the same as 
that taken in the proof of finiteness of the so-called $\beta$-deformed 
$\mathcal{N}=4$ SYM~\cite{RBAKS1,RBAKS2}, and 
the technical part of our proof is also similar to those given there,
though there are a few important differences. 
This is because the $\beta$-deformation  
is also defined by replacing the ordinary product of the original theory 
by a $*$-product which  
shares important properties with the Moyal product.

We emphasise 
that what we show is not merely the finiteness 
of the non-commutative ${\cal N}=4$ SYM: 
the non-commutative SYM is finite, and the finite result 
is continuous with respect to $C$. 
This is achieved because our power counting
procedures ensure the uniform convergence, which is stronger than 
the mere convergence of Feynman integrals.
The finiteness of the non-commutative ${\cal N}=4$ SYM was proved in \cite{RBJackJones}.

We notice that the application of Weinberg's theorem~\cite{RBWeinberg}
in the lightcone gauge involves some subtlety as first pointed out in \cite{RBSmith}.
This point will be discussed later in this paper.

This paper is organised as follows.
In section \ref{RSNontrivial},  
we elaborate on the non-triviality of the
commutative limit, and show that the finiteness of the
commutative version alone does not imply the smoothness
of the commutative limit. 
The proof of our theorem is given in section \ref{RSProof}. 
We conclude with some discussion in section \ref{RSConclusion}.
In an appendix we give an explicit one-loop computation of 
two point functions in the superfield formulation, 
which is indeed continuous in the $C\rightarrow 0$ limit.

%%%%%%%%%%%%%%%%%%%%%%%%%%%%%%%%%%%%%%%%%%%%%%%%%%%%%%%%%%%%%%%%%%%%%%
%%%%%%%%%%%%%%%%%%%%%%%%%%%%%%%%%%%%%%%%%%%%%%%%%%%%%%%%%%%%%%%%%%%%%%
%%%%%%%%%%%%%%%%%%%%%%%%%%%%%%%%%%%%%%%%%%%%%%%%%%%%%%%%%%%%%%%%%%%%%%
\section{Commutative limit and uniform convergence}
\label{RSNontrivial}
%%%%%%%%%%%%%%%%%%%%%%%%%%%%%%%%%%%%%%%%%%%%%%%%%%%%%%%%%%%%%%%%%%%%%%
%%%%%%%%%%%%%%%%%%%%%%%%%%%%%%%%%%%%%%%%%%%%%%%%%%%%%%%%%%%%%%%%%%%%%%
%%%%%%%%%%%%%%%%%%%%%%%%%%%%%%%%%%%%%%%%%%%%%%%%%%%%%%%%%%%%%%%%%%%%%%
In this section, we discuss why the commutative limit is nontrivial, in particular for the 
$\mathcal{N}=4$ SYM. 
We begin by recalling the so-called UV/IR mixing  in generic non-commutative field theories.
In terms of the Feynman rules, 
the effect of the replacement of usual products by the Moyal products 
(\ref{RFMoyal}) simply amounts to 
introduction of phase factors, 
\begin{equation}
e^{-\frac{i}{2} p_\mu C^{\mu \nu} {p'}_\nu} \label{RFMoyalPhase}
\end{equation}
for each vertex, where $p$ and $p'$ are momenta 
associated with the (external or internal) lines connected to the vertex.

It was then found by explicit one-loop calculations
that some Green functions exhibit new types of 
singularities~\cite{Minwalla:1999px, RBHayakawa1, RBHayakawa2}.
These singularities appear only for 
non-planar diagrams.~\footnote{
For planar diagrams, the dependence of the phase factors 
on the loop momenta cancels out,
and hence we have the same Feynman integrals (which may be divergent) 
as the original commutative theory,
multiplied only by over-all phase factors depending on the external momenta.
In \cite{RBJackJones}, an argument is given to show 
the finiteness of the non-commutative $\mN = 4$ SYM 
focusing only on planar diagrams (since contributions from 
non-planar diagrams are finite).
In order to discuss the singularities in the $C\to 0$ limit,
one has to study the non-planar diagrams.  }
Original UV divergences associated with non-planar diagrams are tamed by 
the rapid oscillations introduced by 
the phase factors (\ref{RFMoyalPhase})~\cite{GonzalezArroyo:1982ub}.
One can interpret this as 
an introduction of an effective UV cut-off,
for the internal momenta, of order $\frac{1}{|Ck|}$. Here $k$ is given 
by some linear combination of external momenta which 
may be 
different for different diagrams.
The integral is finite and behaves singularly in the limit $C \to 0$.
Schematically, they behave like 
\begin{equation}
\frac{1}{(C^{\mu \nu} k_\nu)^2}
\label{RFUVIR1}
\end{equation}
for diagrams which are originally quadratically divergent.
For diagrams which are originally logarithmically divergent,
the singular behaviour is,
\begin{equation}
\log{(C^{\mu \nu} k_\nu)^2}.
\label{RFUVIR2}
\end{equation} 
The appearance of these terms is usually called the
``UV/IR mixing'': they arise from originally UV divergent graphs,
and have singular behaviour in the IR, i.e. when the external momenta 
are sent to zero.

These terms are also singular in the commutative limit, 
$C \rightarrow 0$, which is the subject of this paper.
These singular behaviours arise although
the integrands of the Feynman integrals are smooth
with respect to $C$ as in \eqref{RFMoyalPhase}.
Thus, the $C\rightarrow 0$ limit and the integral do not commute.
More precisely, the $C\rightarrow 0$ limit
and the limit in which the upper bound of the momentum integral is taken to infinity
do not commute.  

We will prove that these two limits 
do commute for the case of the $\mN=4$ model.
One might be tempted to think that
because the singularities at $C\rightarrow 0$ originate in the UV 
divergences, UV-finite theories including 4d $\mN=4$ SYM
admit continuous $C \rightarrow 0$  limits.
In fact this reasoning is not sufficient to ensure the continuity. 
The point missed in this argument is that the finiteness 
of the original theory is a result of cancellations
of, say, logarithmically divergent diagrams.
The introduction of the phase factors may tame different 
divergent diagrams in different ways (with different effective cut-offs
for different diagrams). 
If this happens, we will have sums of terms of the form $\log{(Ck)^2}$,
where $k$ can be different for various diagrams.
Such sums can behave singularly 
in the limit $C \rightarrow 0$.  
In other words, the original cancellation is ruined.
In this manner, a theory 
which is finite at $C=0$ can have 
a non-smooth  $C \rightarrow 0$ limit.
Simple one-dimensional integrals (\ref{RFIllustFirst})-(\ref{RFIllustLast})
with this property are presented 
at the end of this section.

Explicit one-loop computations for super Yang-Mills
theories 
with lower supersymmetry
were performed in \cite{RBTravagliniKhoze, RBZanonN12, RBBichlEtal}.
In these computations UV/IR mixing terms are found,
even for some $\mathcal{N}=2$ UV finite theories.
Properties of non-commutative $\mathcal{N}=2$ theories related to
UV/IR mixing are discussed in \cite{RBArmoniMinasianTheisen}.
For the $\mathcal{N}=4$ theory no UV/IR mixing terms are 
found~\cite{RBTravagliniKhoze, RBZanonEtalN4, RBFerrariEtal}
at the one-loop level.

One can look at this problem of the commutative limit 
from a more mathematical point of view.
It is well-known that what guarantees 
the validity of the exchange of two limits is the
condition of uniform convergence;
the cancellation of the divergence at $C=0$ does not imply
a smooth behaviour for $C\rightarrow 0$. 

Let us recall the definition of uniform convergence.
We consider an integral of the form
\begin{equation}
\lim_{\Lambda \rightarrow \infty} \int^{\Lambda} f(p, C) d p
= F(C)
\end{equation}
where $\Lambda$ is the upper bound of the momentum integral~\footnote{
We note that this $\Lambda$ has a slightly different character compared
to the usual UV cutoff in quantum field theory.
Usually one introduces the UV cutoff to make sense out of 
a divergent integral, to define the (perturbation) theory. 
It is a non-trivial issue which type of the cutoff procedure
(such as the simple cut-off, the Pauli-Villars method) one should employ.
For the $\mathcal{N}=4$ SYM,
since the integrals are finite
the type of the cutoff is hardly an issue.
We note that the integrals are made finite by suitable combinations of 
divergent diagrams.}.
The definition of uniform convergence with respect to the parameter $C$ is 
\begin{equation}
{\rm for}\ 
{}^\forall \varepsilon>0, 
\quad
{}^\exists \Lambda_0 
\quad 
{\rm s.t.}
\quad 
\left|\int^{\Lambda} f(p, C) d p - F(C)\right| < \varepsilon
\quad
{\rm for}\ 
{}^\forall \Lambda>\Lambda_0
\ {\rm and}\ 
{}^\forall C,  
\label{uniform_convergence}
\end{equation}
while the definition of the usual convergence (for each fixed value of $C$) is 
given by the condition,
\begin{equation}
{\rm for}\ 
{}^\forall \varepsilon>0, 
\quad
{}^\exists \Lambda_0 
\quad 
{\rm s.t.}
\quad 
\left|\int^{\Lambda} f(p, C) d p - F(C)\right| < \varepsilon
\quad
{\rm for}\ 
{}^\forall \Lambda>\Lambda_0. 
\end{equation}
(Here $\varepsilon$ is the error in the computation of the  
total integral $F(C)$; if one wishes to compute $F(C)$ 
within this error then one has to choose the upper bound 
of the integral, $\L$,
to be larger than the value $\L_0$.
Thus this $\Lambda_0$ can be thought 
of as a quantity which measure the slowness of the convergence; 
larger $\L_0$ implies slower convergence.)
The only difference between the two definitions is 
the extra ``${}^\forall C$'' in the former: 
for the usual convergence, $\L_0$ may depend on 
$C$, whereas for the uniform convergence $\L_0$ does
not depend on $C$. In case of the non-uniform convergence,
$\Lambda_0$ can be singularly large for certain values of $C$,
which makes it possible for $F(C)$ to develop singularities or 
discontinuities at these points even for smooth $f(p,C)$. 
If the condition of the uniform convergence is satisfied,
smoothness properties of the integrand $f(p,C)$ with respect to $C$, 
such as the continuity, 
transfer to that of the $F(C)$. 
In particular, we will use the theorem~\footnote{
See, for example, \cite{Whittaker-Watson}.}
which states that
if $f(p, C)$ is continuous in both $p$ and $C$ and 
if the convergence  is uniform, 
$F(C)$ is also continuous in $C$.

We conclude this section by illustrating
our argument above by some simple
one-dimensional integrals.
We start from the simplest logarithmically divergent integral
\begin{equation}
\int_\mu^\L \frac{dp}{p} =\log\frac{\L}{\mu}. \label{RFIllustFirst}
\end{equation}
Here the parameter $\mu$ plays the role of the IR cutoff. 
Introducing the phase factor, we consider an integral 
\begin{equation}
\int_\mu^\infty \frac{1}{p} e^{i p C k} dp \sim -\log(\mu Ck)
\end{equation}
The right hand side is the leading behaviour 
for $\mu C k \ll 1$,
which can be derived by using a transformation
of the integration variable similar to that presented below.
For definiteness, we assume $C > 0, k > 0$.
The logarithmically divergent integral is now tamed by the oscillating 
phase factor to yield a finite result, which behaves singularly
in the limit $Ck \rightarrow 0$.
This is analogous to 
the ``UV/IR mixing'' for generic non-commutative field theories.

Now let us present an example which yields
a finite result when putting $C=0$ in the integrand,
but nonetheless have a discontinuity in the $C\rightarrow 0$ limit,
\begin{equation}
F(C)=\lim_{\L\rightarrow \infty} \int_\mu^\L \left( \frac{1}{p}e^{ip C k}-\frac{1}{p}e^{ipC k'}
\right) dp.
\label{RFInterestingOnedimensionalIntegral}
\end{equation}
If one puts $C=0$ in the integrand, or equivalently,
if one takes the $C\rightarrow 0$ limit before the 
$\L \rightarrow\infty $ limit,
one gets,
\begin{equation}
F(0)=\int_\mu^\infty\left(\frac{1}{p}-\frac{1}{p}\right)dp=0,
\end{equation}
which is of course finite.
Meanwhile, one can evaluate the integral (\ref{RFInterestingOnedimensionalIntegral})
by using simple
transformations of variables $u= p C k$ and $u=p C k'$, 
\begin{eqnarray}
F(C)=\int_{\mu C k}^{\mu C k'}\frac{1}{u}e^{iu} du \sim \log\left(\frac{k'}{k}\right). 
\label{RFIllustLast}
\end{eqnarray}
Again, the right hand side is the leading behaviour when $C$ is small.
Thus $F(C)$ has a discontinuity at $C=0$ and the $C \to 0$ 
limit is not smooth.
This corresponds to the dangerous situation, where the cancellation between 
would-be divergent terms
are ruined by the phase factors, yielding singular $C\rightarrow 0$ limit. 
We will rule out occurrence of analogous situations 
for Feynman integrals in the $\mN =4$ SYM 
model in the next section.

%%%%%%%%%%%%%%%%%%%%%%%%%%%%%%%%%%%%%%%%%%%%%%%%%%%%%%%%%%%%%%%%%%%%%%
%%%%%%%%%%%%%%%%%%%%%%%%%%%%%%%%%%%%%%%%%%%%%%%%%%%%%%%%%%%%%%%%%%%%%%
%%%%%%%%%%%%%%%%%%%%%%%%%%%%%%%%%%%%%%%%%%%%%%%%%%%%%%%%%%%%%%%%%%%%%%
\section{The proof }
\label{RSProof}
%%%%%%%%%%%%%%%%%%%%%%%%%%%%%%%%%%%%%%%%%%%%%%%%%%%%%%%%%%%%%%%%%%%%%%
%%%%%%%%%%%%%%%%%%%%%%%%%%%%%%%%%%%%%%%%%%%%%%%%%%%%%%%%%%%%%%%%%%%%%%
%%%%%%%%%%%%%%%%%%%%%%%%%%%%%%%%%%%%%%%%%%%%%%%%%%%%%%%%%%%%%%%%%%%%%%
Our proof is technically similar to those given in \cite{RBAKS1,RBAKS2},
and in \cite{RBBrinkLindgrenNilssonProof},
and our main focus will be on differences
in particular on the manner uniform convergence is achieved. 
We will be only concerned with the UV properties of the Feynman integral.
We shall assume below that there is an implicit IR cut-off to 
avoid any possible complication from 
IR divergences.
We note that the phase factor associated with the Moyal product 
does not introduce the rapid oscillations
in the IR (when the loop momenta are small), and hence it seems likely that
the structure of IR divergences is not affected much by the non-commutativity.

The outline of the proof is as follows.
In section \ref{RSSLCsuperspace}, we formulate the non-commutative $\mathcal{N}=4$
SYM in terms of the lightcone superfield. 
Due to properties of the Moyal product
such as the associativity,
the result is quite simple:
one replaces products between superfields in the 
superspace action for the commutative theory with Moyal products.

In section \ref{RSSPowercounting}, we evaluate (an upper bound for) 
the superficial degree of divergence $D$ of 
Feynman integrals. This is done in two steps.
In the first step, we make a ``rough estimate'' of $D$,
by using techniques of evaluating superfield Feynman graphs
similar to those introduced in \cite{RBGrisaruSiegelRocek} for 
$\mathcal{N}=1$ supergraphs. At this stage one concludes that $D\sim 0$.
In the next step, one focusses on vertices connected to
external lines; using the particular form of the vertices,
one can improve the rough estimate of $D$ to show that $D$ is in fact negative.~\footnote{
We note that in the lightcone gauge there is no wavefunction renormalisation.
In some gauge there is wavefunction renormalisation 
which does not affect physical observables of the theory.}

Finally, we use Weinberg's theorem~\cite{RBWeinberg} in section \ref{RSSWeinbergTh}.
In our context Weinberg's theorem implies the uniform convergence,
which in turn results in the continuity of the result of the Feynman integrals
with respect to $C$.

%%%%%%%%%%%%%%%%%%%%%%%%%%%%%%%%%%%%%%%%%%%%%%%%%%%%%%%%%%%%%%%%%%%%%%
%%%%%%%%%%%%%%%%%%%%%%%%%%%%%%%%%%%%%%%%%%%%%%%%%%%%%%%%%%%%%%%%%%%%%%
%%%%%%%%%%%%%%%%%%%%%%%%%%%%%%%%%%%%%%%%%%%%%%%%%%%%%%%%%%%%%%%%%%%%%%
\subsection{Non-commutative $\mN=4$ SYM in lightcone superspace}
\label{RSSLCsuperspace}
\hspace{0.51cm}
%%%%%%%%%%%%%%%%%%%%%%%%%%%%%%%%%%%%%%%%%%%%%%%%%%%%%%%%%%%%%%%%%%%%%%
%%%%%%%%%%%%%%%%%%%%%%%%%%%%%%%%%%%%%%%%%%%%%%%%%%%%%%%%%%%%%%%%%%%%%%
%%%%%%%%%%%%%%%%%%%%%%%%%%%%%%%%%%%%%%%%%%%%%%%%%%%%%%%%%%%%%%%%%%%%%%
In this section we introduce the lightcone superfield
formalism in the non-commutative space,
which is a natural  extension of 
the original formulation in the commutative space \cite{RBBLNAction}. 

We define the lightcone coordinates
\begin{eqnarray}
x^{\pm} = \frac{x^0 \pm x^3}{\sqrt{2}},  
\end{eqnarray} 
where $x^+$ plays the role of the time coordinate.
The remaining  two coordinates $x^1, x^2$ are non-commutative,
\begin{equation}
x^1\star x^2 - x^2\star x^1 = iC^{12}=i C.
\end{equation}
Our metric convention is $\eta^{\mu\nu}=diag(-1,+1,+1,+1)$,
and the lightcone components of the metric are
$\eta_{+-}=\eta_{-+}=\eta^{+-}=\eta^{-+}=-1, 
\eta_{++}=\eta_{--}=\eta^{++}=\eta_{--}=0$. 
We use indices $\m, \n=0, \ldots, 3$ for spacetime coordinates.
The lightcone components of the gauge fields are  
$ A_{\pm} = (A_0 \pm A_3)/\sqrt{2}$. 
We impose the lightcone gauge condition 
\begin{eqnarray}
A_-=0. 
\end{eqnarray}
In this gauge, $A_\pm$ are not propagating. 

There are eight bosonic propagating degrees of freedom: 
two transverse components of the gauge field $A^1$ and $A^2$, 
which we combine into a complex field 
$A=\frac{1}{\sqrt{2}}(A^1+iA^2)$ and $\bar{A} = \frac{1}{\sqrt{2}}(A^1-iA^2)$
and three complex scalar fields, $\varphi_{mn}=-\varphi_{nm} (m,n=1,\cdots,4)$ 
with the condition $\bar{\varphi}_{mn}=\epsilon_{mnpq}\varphi^{pq}/2$, 
where
$\epsilon_{mnpq}$ is a totally antisymmetric tensor with $\epsilon_{1234}=+1$.   
Half of the spinor fields are not propagating in the lightcone gauge, 
and there are four complex (single-component) fermions $\chi^m(m=1,\cdots,4)$. 

The action in the lightcone gauge is obtained from the original
action by eliminating non-dynamical degrees of freedom such as $A_+$~\cite{RBBLNAction}.
In this procedure the trace cyclicity of the matrix product plays an essential role.
The procedure goes through similarly in the non-commutative case, 
because the Moyal product 
also 
satisfies the cyclicity inside the trace\footnote{
This may be understood as a consequence of 
the mapping between functions in the non-commutative space and matrices~\cite{Aoki:1999vr}: 
the integral and the non-commutative product are
identified to the trace and the product of matrices, 
respectively. 
}, 
\begin{eqnarray} 
\int \left( f_1\star f_2\star\cdots\star f_n\right) d^dx
=
\int  \left(f_2\star\cdots\star f_n\star f_1 \right)d^dx. 
\end{eqnarray} 
The invariance under supersymmetry is also preserved in a similar way.

Now let us introduce the superfield formulation.
There are four bosonic and eight fermionic coordinates, 
$x^+,x^-, z=(x^1+i x^2)/\sqrt{2}, \bar{z}=(x^1-ix^2)/\sqrt{2}$
and $\theta^m,\bar{\theta}_m (m=1,\cdots,4)$. 
Eight kinematical (manifest) supersymmetries are generated by 
\begin{eqnarray}
Q^m
=
-\frac{\partial}{\partial\bar{\theta}_m}
-
\frac{i}{\sqrt{2}}\theta^m\partial_-, 
\qquad
\bar{Q}_m
=
\frac{\partial}{\partial\theta^m}
+
\frac{i}{\sqrt{2}}\bar{\theta}_m\partial_-. 
\end{eqnarray}
The superspace chiral derivatives are defined by 
\begin{eqnarray}
d^m
=
-\frac{\partial}{\partial\bar{\theta}_m}
+
\frac{i}{\sqrt{2}}\theta^m\partial_-, 
\qquad
\bar{d}_m
=
\frac{\partial}{\partial\theta^m}
-
\frac{i}{\sqrt{2}}\bar{\theta}_m\partial_-. 
\end{eqnarray}
The scalar superfield $\Phi$ and its hermitian conjugate $\bar{\Phi}$ 
satisfy the chirality condition 
\begin{eqnarray}
d^m\Phi=0, 
\qquad
\bar{d}_m\bar{\Phi}=0
\label{chiral_superfield_1}
\end{eqnarray}
and
\begin{equation}
\bar{\Phi} =\frac{\bar{d}^4}{2 \partial_-^2} \Phi, \qquad
\Phi =\frac{d^4}{2 \partial_-^2} \bar{\Phi}
\label{RFphibarintermsofphi}
\end{equation}
We use abbreviations such as, 
\begin{equation}
\bar{d}^4=
\bar{d}_1
\bar{d}_2
\bar{d}_3
\bar{d}_4
=\frac{1}{24}
\epsilon^{mnpq}
\bar{d}_m
\bar{d}_n
\bar{d}_p
\bar{d}_q.
\end{equation}
This convention differs from those of \cite{RBBLNAction, RBBrinkLindgrenNilssonProof,
RBAKS1, RBAKS2}
by a factor of $24$.
The definition of the superfields remains the same for the non-commutative case 
since 
derivatives in the lightcone coordinates commute with $\star$-products.

In terms of the component fields, the scalar superfield $\Phi$ is expressed as 
\begin{eqnarray}
\Phi(x,\theta,\bar{\theta})
&=&
-\frac{1}{\partial_-}A(X)
-\frac{i}{\partial_-}\theta^m\bar{\chi}_m(X)
+\frac{i}{\sqrt{2}}\theta^m\theta^n\bar{\varphi}_{mn}(X)
\nonumber\\
& &
+
\frac{\sqrt{2}}{6}\epsilon_{mnpq}\theta^m\theta^n\theta^p\chi^q(X)
-
\frac{1}{12}\epsilon_{mnpq}\theta^m\theta^n\theta^p\theta^q\partial_-\bar{A}(X).
\end{eqnarray}
Here $X$ is the chiral coordinate 
$X=(x^+,y^-, z, \bar{z})$  where $y^-
\equiv x^--\frac{i}{\sqrt{2}}\theta^m\bar{\theta}_m$.

The action in terms of superfields is~\footnote{
We use the prescription by Mandelstam~\cite{RBMandelstam} when defining
factors such as $\frac{1}{\partial_-}$, which enables us to 
perform the Wick rotation.
}${}^,$\footnote{
We follow the notation used in \cite{RBBLNAction}, 
$\partial=(\partial^1+i\partial^2)/\sqrt{2}, 
\bar{\partial}=(\partial^1-i\partial^2)/\sqrt{2} 
$. }
\begin{eqnarray}
S
=
\frac{1}{8}\int d^4x\int d^4\theta d^4\bar{\theta}\ \Tr
\Biggl\{
-2\bar{\Phi}\frac{\square}{\partial_-^2}\Phi
+
\frac{8ig}{3}\left(
\frac{1}\partial_-{}\bar{\Phi}\cdot [\Phi,\bar{\partial}\Phi]_\star
+
\frac{1}\partial_-{}\Phi\cdot [\bar{\Phi},\partial\bar{\Phi}]_\star
\right)
\nonumber\\
+
2g^2\left(
\frac{1}{\partial_-}[\Phi,\partial_-\Phi]_\star
\cdot 
\frac{1}{\partial_-}[\bar{\Phi},\partial_-\bar{\Phi}]_\star
+
\frac{1}{2}[\Phi,\bar{\Phi}]_\star[\Phi,\bar{\Phi}]_\star
\right)
\Biggl\}
\label{RFSuperspaceAction}
\end{eqnarray}
where the star commutator between two fields $A$, $B$ is defined by
\begin{equation}
[A,B]_\star=A\star B-B\star A.
\end{equation}
The action of the non-commutative SYM is the same as the original theory
in the commutative space, except for the replacement of the product with
the Moyal product.
This is similar to the formulation of $\beta$-deformed $\mathcal{N}=4$ SYM
in terms of the lightcone superfield discussed in \cite{RBAKS1, RBAKS2}.
  
%%%%%%%%%%%%%%%%%%%%%%%%%%%%%%%%%%%%%%%%%%%%%%%%%%%%%%%%%%%%%%%%%%%%%%
%%%%%%%%%%%%%%%%%%%%%%%%%%%%%%%%%%%%%%%%%%%%%%%%%%%%%%%%%%%%%%%%%%%%%%
%%%%%%%%%%%%%%%%%%%%%%%%%%%%%%%%%%%%%%%%%%%%%%%%%%%%%%%%%%%%%%%%%%%%%%
\subsection{Power counting}
\label{RSSPowercounting}
\hspace{0.51cm}
%%%%%%%%%%%%%%%%%%%%%%%%%%%%%%%%%%%%%%%%%%%%%%%%%%%%%%%%%%%%%%%%%%%%%%
%%%%%%%%%%%%%%%%%%%%%%%%%%%%%%%%%%%%%%%%%%%%%%%%%%%%%%%%%%%%%%%%%%%%%%
%%%%%%%%%%%%%%%%%%%%%%%%%%%%%%%%%%%%%%%%%%%%%%%%%%%%%%%%%%%%%%%%%%%%%%

In this section, we consider 
the superficial degree of divergence $D$. 
In usual field theory, $D$ is determined by counting the powers of momenta.
In the non-commutative theory the integrand depends on the
momenta non-polynomially due to the phase factors introduced by the $\star$-product.
We define $D$ neglecting  
the phase factors.
The superficial degree of divergence so defined is useful
when we apply Weinberg's theorem as we will see in 
section \ref{RSSWeinbergTh}.

The power counting procedure is  divided into two steps. 
In the first step a ``rough'' estimate of $D$ is made, which is refined in the second step.
The starting point of the first step is to write down 
the superspace Feynman rules. The propagator is given by
\begin{eqnarray}
&&
\langle 
\Phi_{p_{(1)}}{}^u{}_v \left(\th_{(1)}, \bar\th_{(1)}\right)
\Phi_{p_{(2)}}{}^r{}_s \left(\th_{(2)}, \bar\th_{(2)}\right)
\rangle
\nonumber\\
&=&
\frac{1}{(2\pi)^4}\delta^4(p_{(1)}+p_{(2)})\times
\delta^u{}_s \delta^r{}_v \times
\frac{i}{p_\m^2}
d^4_{(1)}
\delta^8(\th_{(1)}-\th_{(2)}),
\label{RFSuperPropagator}
\end{eqnarray}
where
\begin{eqnarray}
\delta^8\left(\theta-\theta'\right)
=
\left(\theta-\theta'\right)^4
\left(\bar{\theta}-\bar{\theta}'\right)^4.  
\end{eqnarray} 
Our convention is 
$(\theta)^4 =(1/24) \epsilon_{mnpq}\th^m\th^n\th^p\th^q$.
The vertices can be read off from the action 
by using the formula (\ref{RFphibarintermsofphi}).

We will now sketch the supergraph power-counting. For details, see 
\cite{RBBrinkLindgrenNilssonProof, RBAKS1, RBGrisaruSiegelRocek, RBThousandonelessons}.
When evaluating a Feynman diagram, one 
first performs the $\th$-integrals. Focussing on a single internal line, 
one can get rid of $d$'s originating from
the propagator and $\bar{d}$'s originating from vertices by using 
partial integration, ending up with a bare superspace $\delta$-function.
Then the $\th$-integral can be performed, eliminating one $\th$ variable. 
This procedure
is to be repeated to the point where only one $\th$-integral is left.
In this process, for each loop, one has to use the following identity once:
\begin{eqnarray}
\delta^8\left(\theta_{(1)}-\theta_{(2)}\right)
d_{(1)}^4\bar{d}_{(1)}^4
\delta^8\left(\theta_{(1)}-\theta_{(2)}\right)
=
\delta^8\left(\theta_{(1)}-\theta_{(2)}\right). 
\label{RFGSR}
\end{eqnarray} 
Other combinations of two $\delta$-functions and
chiral derivatives vanish under the 
$\theta$-integral~\cite{RBGrisaruSiegelRocek, RBBrinkLindgrenNilssonProof}.
This means that we lose $4$ powers of momentum\footnote{We note that $d$'s or $\bar d$'s
should be thought
of as a square root of momenta, in the power counting procedure.} for each loop.
This cancels the original $4$ powers of momentum from the loop integral.
Thus, the contribution of each loop to the superficial degree of divergence $D$ is zero.
\footnote{
For the $\beta$-deformed theory~\cite{RBAKS1, RBAKS2}
the equality \eqref{RFGSR} is modified
except for planar diagrams,
because the $*$-product for the $\beta$-deformation acts
on the $\theta$-space.
Hence the analysis was restricted to
the planar level for the $\beta$-deformed theory.
In the present case, this equality remain unchanged, because the Moyal product 
does not act on the $\th$'s, 
so that our analysis is valid for all diagrams including non-planar ones.
}
The contribution of the propagator to $D$ comes from the $\frac{d^4}{p^2}$ part of 
(\ref{RFSuperPropagator})
and is zero.
The contributions of the vertices are also zero as can be read off from the action
(\ref{RFSuperspaceAction}). 

The result of the first step in the power counting procedure is thus
$D\sim 0$. At this stage,  %In deriving the conclusion 
we are not distinguishing 
the external and internal momenta. In the second step, we  
distinguish them,  focussing 
on a vertex attached to an external line. By certain manipulations using 
the explicit form of the vertices one can then show that the 
superficial degree of divergence decreases
by one (or more). These manipulations are, (a) 
moving $d$'s or $\bar{d}$'s
from internal lines to external lines
via partial integrations 
(b) cancellations between contributions from different vertices and contractions,
in the leading behaviour when the internal momenta are much larger than
the external momenta.
One has to do this analysis for all possible contractions 
of all three-point and four-point vertices.
Here, one has to verify that the cancellations used in this step occur among 
contractions which acquires the same phase factors from the non-commutativity.

\begin{figure}
\begin{center}
\scalebox{0.3}{
\includegraphics{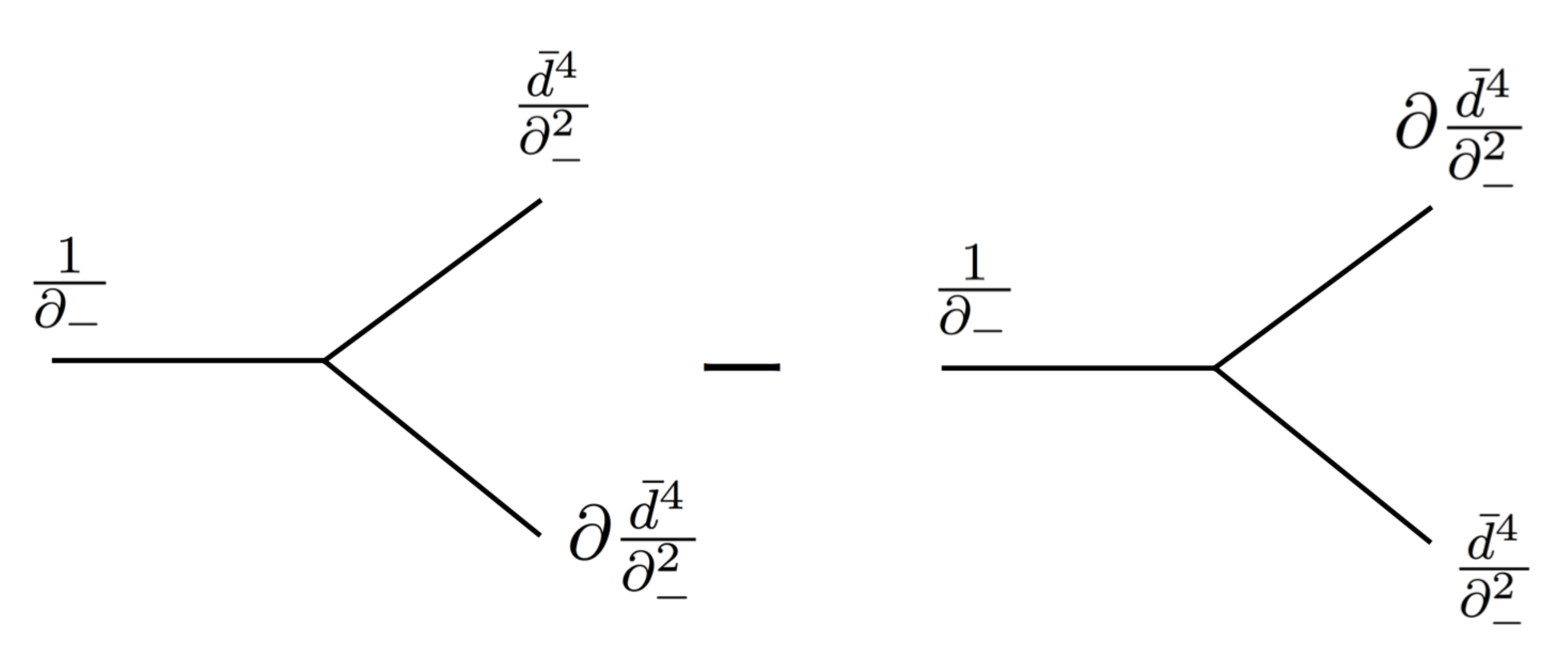}}
\end{center}
\caption{Three vertex \eqref{eq:3pt_vertex}.}\label{RP3Vertex}
\end{figure}

\begin{figure}
\begin{center}
\includegraphics{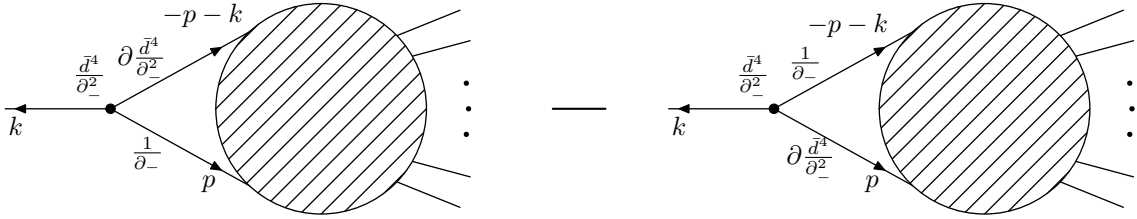}
\end{center}
\caption{
A class of generic processes in which the vertex \eqref{eq:3pt_vertex} is involved.
The cancellation between two sets of diagrams 
takes place as shown in \eqref{eq:cancellation_3pt}. 
}
\label{RPContractionExample}
\end{figure}

This step is 
parallel to the corresponding step in the proof of finiteness
of the $\beta$-deformed theory given in \cite{RBAKS1},
and we will not give the details in this paper. 
The present case is actually simpler since the chiral derivatives commute with 
the Moyal products.  
In order to illustrate the procedure let us discuss one particular example
of the arguments used in this step, for the three-point vertex,
\begin{eqnarray}
\frac{ig}{3}
\int d^4x\int d^4\theta d^4\bar{\theta}\ \Tr
\left(
\frac{1}\partial_-{}\Phi\cdot [\bar{\Phi},\partial\bar{\Phi}]_\star
\right)
=
\frac{ig}{12}
\int d^4x\int d^4\theta d^4\bar{\theta}\ \Tr
\left(
\frac{1}\partial_-{}\Phi\cdot \left[
\frac{\bar{d}^4}{\partial_-^2}\Phi,
\partial
\frac{\bar{d}^4}{\partial_-^2}\Phi
\right]_\star
\right),
\label{eq:3pt_vertex}
\end{eqnarray} 
which can be represented diagrammatically as in Fig.~\ref{RP3Vertex}.
In our convention products of fields are always taken to be counter-clockwise in 
Feynman diagrams.
The contributions we consider are shown in Fig.~\ref{RPContractionExample}.
The 
shaded disk represents general processes.
We are focussing on a particular vertex connected to the external line,
which is given by Fig.~\ref{RP3Vertex}. 
The contractions we consider are shown
in Fig.~\ref{RPContractionExample}. 
By moving  $\bar{d}^4$ appropriately, it is possible to show 
that the contributions from two diagrams
are the same except for the momentum factor.
The sum of these momentum factors are
\begin{equation}
- \frac{p+k}{k_-^2 p_- (p+k)_-^2} +
\frac{p}{k_-^2 p_-^2 (p_-+k_-)}.
\label{eq:cancellation_3pt}
\end{equation}
The leading terms for $p \gg k$ cancel out. 
This cancellation implies that the superficial degree of divergence is decreased by one, improving the convergence.
It is essential that the cancellation
is not affected by the phase factors 
originating in the Moyal product; the phase factors associated with 
the two vertices shown in Fig. \ref{RPContractionExample} are identical.

As discussed in \cite{RBAKS1}, there are a few exceptional diagrams 
in which general arguments do not apply. They are one-loop diagrams and
are evaluated explicitly in the appendix.

%%%%%%%%%%%%%%%%%%%%%%%%%%%%%%%%%%%%%%%%%%%%%%%%%%%%%%%%%%%%%%%%%%%%%%
%%%%%%%%%%%%%%%%%%%%%%%%%%%%%%%%%%%%%%%%%%%%%%%%%%%%%%%%%%%%%%%%%%%%%%
%%%%%%%%%%%%%%%%%%%%%%%%%%%%%%%%%%%%%%%%%%%%%%%%%%%%%%%%%%%%%%%%%%%%%%
\subsection{Weinberg's theorem, UV finiteness and the commutative limit}
\label{RSSWeinbergTh}
\hspace{0.51cm}
%%%%%%%%%%%%%%%%%%%%%%%%%%%%%%%%%%%%%%%%%%%%%%%%%%%%%%%%%%%%%%%%%%%%%%
%%%%%%%%%%%%%%%%%%%%%%%%%%%%%%%%%%%%%%%%%%%%%%%%%%%%%%%%%%%%%%%%%%%%%%
%%%%%%%%%%%%%%%%%%%%%%%%%%%%%%%%%%%%%%%%%%%%%%%%%%%%%%%%%%%%%%%%%%%%%%
At each order in 
perturbation theory,
we have finite sums of terms of the form 
\begin{equation}
\int f(p,C) dp= \int g(p) 
e^{i \sum_{( i )} p_{( i )\mu} C^{\mu\nu} p'_{( i )_\nu}}
dp.
\label{RFIntegralInTermsOfRationalPartAndPhasePart}
\end{equation}
The integrand is given by a
rational function $g(p)$ multiplied by a single phase factor.
The arguments in the previous section shows that
the superficial degree divergence of $\int g(p) dp$ is negative.
(We recall that our definition of $D$ does not include the phase factor.)
Here $p_{(i)}$, and $p'_{(i)}$ are some linear combinations of the internal
and external momenta.

It is also easy to see that the same holds for all sub-diagrams.
One can now invoke Weinberg's theorem~\footnote{
The assumption of Weinberg's theorem is that the superficial degree of divergence
is negative for all possible linear subspaces in the integration variables~\cite{RBWeinberg}.
For Lorentz invariant Feynman integrals (after Wick rotation), the denominator 
depends on the momenta always in the form $p_\mu^2$, and therefore
it is sufficient to consider all possible subgraphs to 
guarantee that this requirement is met.
In the lightcone gauge, there are factors of $1/\partial_-$ in 
the integrand. 
As a consequence, it is necessary to separately examine the linear subspaces
distinguishing longitudinal and transverse components
for all loop momenta. 
For example, one should consider the region in which 
transverse components are sent to infinity but
longitudinal components are kept finite.
This was not done in the original finiteness proof  
of 4d $\mathcal{N}=4$ 
theory in \cite{RBBrinkLindgrenNilssonProof}.
(For the proof given in \cite{RBMandelstam}, this 
subtlety is pointed out and a resolution of it is discussed in \cite{RBSmith}.)
We will make a few comments on this subtlety also in section \ref{RSConclusion}.
}, which 
assures the absolute convergence of the integral $\int g(p) dp$, i.e.  the 
convergence of the integral 
\begin{equation}
\int |g(p)| dp. \label{RFIntegralAbsolute}
\end{equation}

A quick way to see that 
\eqref{RFIntegralAbsolute} implies the uniform convergence of  (\ref{RFIntegralInTermsOfRationalPartAndPhasePart}) is the following.
The convergence of (\ref{RFIntegralAbsolute}), or
the absolute convergence of
(\ref{RFIntegralInTermsOfRationalPartAndPhasePart}),
guarantees that the original integral (\ref{RFIntegralInTermsOfRationalPartAndPhasePart}) converges to a definite value, $F(C)$.
Then,
$\left|\int^{\Lambda} f(p, C) d p - F(C)\right|$ in the condition 
for uniform convergence, \eqref{uniform_convergence}, 
can be rewritten as\footnote{
We use the notation $\int^\infty_\L$ 
to denote the (multi-dimensional) integral 
complementary to $\int^\L$, i.e.~$\int^\infty=\int^\L+ \int_\Lambda^\infty$. 
}
\begin{equation}
\left|\int^\infty_\L f(p,C) dp \right|
=
\left|\int^\infty_\L g(p)e^{i\sum pCp'} dp \right|,
\end{equation}
and the r.h.s. satisfies the elementary inequality
\begin{equation}
\left| 
\int^\infty_\L g(p)e^{i\sum pCp'} dp 
\right|
< \int^\infty_\L |g(p)|dp.
\end{equation}
Because of the convergence of (\ref{RFIntegralAbsolute}),
for arbitrary $\varepsilon >0$ there exists $\Lambda_0$ 
such that for any $\Lambda > \Lambda_0$ 
$
\int^\infty_\L |g(p)|dp <\varepsilon
$.
By using the same $\varepsilon$ and $\Lambda$
we have
\begin{equation}
\left|\int^{\Lambda} f(p, C) d p - F(C)\right|
=
\left|\int^\infty_\L f(p,C) dp \right| <\varepsilon
\end{equation}
for arbitrary $C$, which is the condition of uniform convergence.

We now use the theorem which states 
that if $f(p, C)$ is continuous in $p$ and $C$ 
and $\L \rightarrow \infty$ is uniformly convergent,
$F(C)$ is continuous in $C$~\cite{Whittaker-Watson}.
Thus we have shown that there is no discontinuity in $C$,
in particular for $C\rightarrow 0$.

%%%%%%%%%%%%%%%%%%%%%%%%%%%%%%%%
%%%%%%%%%%%%%%%%%%%%%%%%%%%%%%%%
%%%%%%%%%%%%%%%%%%%%%%%%%%%%%%%%
\section{Conclusion and discussion}\label{RSConclusion}
\hspace{0.51cm}
%%%%%%%%%%%%%%%%%%%%%%%%%%%%%%%%
%%%%%%%%%%%%%%%%%%%%%%%%%%%%%%%%
%%%%%%%%%%%%%%%%%%%%%%%%%%%%%%%%

In this paper we have analysed the UV properties of
the non-commutative version of the 4d $\mN=4$ SYM. 
We have shown that the cancellations between diagrams at $C=0$ 
i.e. of the commutative theory, which are responsible for the finiteness
of the commutative theory, persist for the non-commutative theory as well.
These cancellations ensure that the momentum integrals
converge uniformly with respect to $C$, which in turn implies
that the Green functions (in the lightcone gauge) have no discontinuity in $C$,
to all order in perturbation theory.

This continuity is one of the key steps of the non-perturbative definition 
of 4d $\mN=4$ SYM proposed in \cite{Hanada:2010kt}.
(The proposal for 3d maximal SYM~\cite{Maldacena:2002rb}
also includes implicitly the assumption that the commutative limit is continuous.
In this case, however, 
our proof does not directly apply because 
we utilised the independence of lightcone coordinates and the Moyal product;
in three dimension, 
in the presence of the
non-commutativity in two space-like directions, 
the lightcone coordinates become non-commutative inevitably.) 
The proposal may eventually enable 
us to study nonperturbative features of 4d $\mN=4$ SYM numerically, 
which should deepen our understanding 
of the AdS/CFT correspondence and may make it possible to 
study quantum aspects of gravity from 
a dual gauge theory. Such numerical approach has been
so far successful for the $(0+1)$-d theory 
\cite{RBBFSS,RBdWHN}\cite{Itzhaki:1998dd} (for recent work see \cite{Hanada:2013rga}) and $(1+1)$-d theory \cite{Catterall:2010fx}. 
The 4d theory, which has been considered much more extensively in the past, 
would serve as an even better laboratory. 

The non-commutative 4d $\mathcal{N}=4$ SYM is also an interesting 
theory in its own right. It is believed that this theory has a gravity dual,
and aspects of the duality have been studied for example in
\cite{RBHashimotoItzhaki, RBMaldacenaRusso, RBMatsumotoYoshida}. 

The appearance of singularities of the form (\ref{RFUVIR1}) and (\ref{RFUVIR2}),
is a characteristic feature of  non-commutative theories.~\footnote{
Usually, the singular terms (\ref{RFUVIR1}) and (\ref{RFUVIR2}) 
are characterised by the singular behaviour in the IR region of 
the external momenta.
What we have studied in this paper 
is the smoothness properties when the non-commutative parameter goes to zero.
Our analysis alone does not exclude the occurrence of IR singularities, since 
it is at least logically possible to have a term which behaves singularly 
when external momenta go to zero but does not have a discontinuity in $C$.
}
These singularities play important roles in the study of the relations between 
non-commutative field theories and
ordinary field theories.~\footnote{
For a recent interesting discussion on the relation between
the analog of the UV/IR mixing effect on a non-commutative sphere
and appearance of a certain non-local
interaction in the renormalisation group flow, 
see \cite{RBKawamotoKurokiTomino}. See also \cite{RBVaidya, RBChuMadoreSteinacker}.
}
They may also have some relevance in some proposals
discussing relation between non-commutative field theories 
and gravity \cite{Steinacker:2010rh,Aschieri:2005zs}.
We stress that our work is the first to establish 
strong constraints on these singularities to the all order in perturbation theory.
It would be interesting to consider more general theories and properties 
from this approach.

We believe that our analysis of the continuity of Green functions
with respect to the non-commutative parameter $C$ 
will be the core in the study of the commutative limit in
the $\mN = 4$ SYM. 
There are directions in which one can extend our analysis
in this paper.
First, 
one can also study 
the differentiability of 
Green functions with respect to $C$
by studying
the convergence property of the integral
which have the integrand given by the 
$C$-derivative of the original integrand. 
Each $C$-derivative acting on the phase factor 
$e^{ipC p'}$ brings in extra two powers of momenta,
increasing the superficial degree of convergence.
It should be possible to clarify the structure
of singularities in the $C$-derivatives of Green functions
by appropriate extension of our method.
Second, we have confined ourselves to study of Green functions
of the fundamental fields. It would be also interesting to study 
gauge invariant operators.
Recently correlation function of composite operators have been studied
in the lightcone gauge formalism~\cite{RBAnanthKovacsParikh}.
Third, in this paper we have used the lightcone gauge. Hence, 
the Lorentz invariance of theory is not manifest. It is natural to expect that
the lightcone gauge formulation is equivalent to a covariant formulation, say,
in the Lorentz gauge, which leads immediately also to the Lorentz invariance.
Although the gauge independence of perturbative gauge theories
is fairly well-established, this issue should be more non-trivial
for non-commutative gauge theories because of the non-locality introduced by Moyal 
products. This issue is not unrelated to the issue of the gauge invariant operators.
When discussing the gauge independence, one has to fix one's attention on a set of
gauge invariant observables. In the standard non-conformal gauge theories, one usually studies 
the S-matrix. Since the (commutative) $\mathcal{N}=4$ theory is conformal, 
a natural candidate for the set of observables are $n$-point correlation functions of
composite operators (with definite conformal dimensions).

In this paper, 
the convergence of the Feynman integrals was studied applying 
Weinberg's theorem, following previous work \cite{RBBrinkLindgrenNilssonProof}. 
In the lightcone superfield formalism Feynman rules are not manifestly
Lorentz invariance in particular because of the appearance of factors 
of $\frac{1}{\partial_-}$.
Hence one should examine
the limit which breaks the symmetry, for example, a regime in which 
transverse components are taken to be large
while longitudinal components of momenta are kept finite.
In appendix \ref{RSAppendix} we will show explicitly 
that there is a more non-trivial cancellation at the one-loop level.
It is possible to classify the UV regions for general Feynman diagrams  and
study the superficial degree of divergence
by using appropriate diagrammatic techniques.
This issue will be addressed in a separate publication.

We hope that this work can  provide a basis for studies of
the application of non-commutative $\mN=4$ SYM and
serve as a starting point to clarify the relation between 
non-commutative field theories and their commutative counterparts
more generally.

%%%%%%%%%%%%%%%%%%%%%%%%%%%%%%%%%%%%%%%%%%%%%%%%%%%%%%%%%%%%%%%%%%%%%%
%%%%%%%%%%%%%%%%%%%%%%%%%%%%%%%%%%%%%%%%%%%%%%%%%%%%%%%%%%%%%%%%%%%%%%
%%%%%%%%%%%%%%%%%%%%%%%%%%%%%%%%%%%%%%%%%%%%%%%%%%%%%%%%%%%%%%%%%%%%%%
\section*{Acknowledgements}
\hspace{0.51cm}
%%%%%%%%%%%%%%%%%%%%%%%%%%%%%%%%%%%%%%%%%%%%%%%%%%%%%%%%%%%%%%%%%%%%%%
%%%%%%%%%%%%%%%%%%%%%%%%%%%%%%%%%%%%%%%%%%%%%%%%%%%%%%%%%%%%%%%%%%%%%%
%%%%%%%%%%%%%%%%%%%%%%%%%%%%%%%%%%%%%%%%%%%%%%%%%%%%%%%%%%%%%%%%%%%%%% 
The authors would like to thank 
S.~Kovacs, S.~Matsuura, F.~Sugino for their careful 
reading of our manuscript and very useful discussions and comments. 
We would like to thank 
S.~Ananth, L.~Brink, L.~Dixon, T.~McLoughlin, and
J.~Nishimura for very useful discussions and comments.  
The work of M.~H. is supported in part by the Grant-in-Aid of the Japanese Ministry 
of Education, Sciences and Technology, Sports and Culture (MEXT) 
for Scientific Research (No. 25287046), and  
the National Science Foundation under Grant No. PHYS-1066293 
and the hospitality of the Aspen Center for Physics. 
%%%%%%%%%%%%%%%%%%%%%%%%%%%%%%%%%%%%%%%%%%%%%%%%%%%%%%%%%%%%%%%%%%%%%%
%%%%%%%%%%%%%%%%%%%%%%%%%%%%%%%%%%%%%%%%%%%%%%%%%%%%%%%%%%%%%%%%%%%%%%
%%%%%%%%%%%%%%%%%%%%%%%%%%%%%%%%%%%%%%%%%%%%%%%%%%%%%%%%%%%%%%%%%%%%%% 
\appendix 
%%%%%%%%%%%%%%%%%%%%%%%%%%%%%%%%%%%%%%%%%%%%%%%%%%%%%%%%%%%%%%%%%%%%%%
%%%%%%%%%%%%%%%%%%%%%%%%%%%%%%%%%%%%%%%%%%%%%%%%%%%%%%%%%%%%%%%%%%%%%%
%%%%%%%%%%%%%%%%%%%%%%%%%%%%%%%%%%%%%%%%%%%%%%%%%%%%%%%%%%%%%%%%%%%%%% 

\section{Explicit one-loop computations}
\label{RSAppendix}
In this appendix we perform explicit one-loop computations
of two-point Green functions of the non-commutative $\mathcal{N}=4$ SYM
in the lightcone superfield formalism.
The asymmetric asymptotic region --
in which the transverse components
goes to infinity while longitudinal components remain finite --
is important.
There are two superfield diagrams. 
Superficial degree of divergence for each diagram
is negative in the usual power counting. However,
the diagrams are both logarithmically divergent 
because of the asymmetric asymptotic region.
The divergent contributions from the two diagrams
cancel each other and the result behaves well in the 
$C\rightarrow 0$ limit.

Below in section \ref{RSSFeynmanRule} we explain the Feynman rules.
The next section \ref{RSSOneLoop} summarises the result for each diagrams 
and explains the cancellations.
In section \ref{RSSNiceFormulae} we compile some useful formulae.

%%%%%%%%%%%%%%%%%%%%%%%%%%%%%%%%%%%%%%%%%%%%%%%%%%%%%%%
%%%%%%%%%%%%%%%%%%%%%%%%%%%%%%%%%%%%%%%%%%%%%%%%%%%%%%%
%%%%%%%%%%%%%%%%%%%%%%%%%%%%%%%%%%%%%%%%%%%%%%%%%%%%%%%
\subsection{Feynman rules for $\mathcal{N}=4$ SYM}
\label{RSSFeynmanRule}
%%%%%%%%%%%%%%%%%%%%%%%%%%%%%%%%%%%%%%%%%%%%%%%%%%%%%%%
%%%%%%%%%%%%%%%%%%%%%%%%%%%%%%%%%%%%%%%%%%%%%%%%%%%%%%%
%%%%%%%%%%%%%%%%%%%%%%%%%%%%%%%%%%%%%%%%%%%%%%%%%%%%%%%
Because $\Phi$ and $\bar\Phi$ are related by \eqref{RFphibarintermsofphi}, 
we can rewrite the action \eqref{RFSuperspaceAction} 
in terms only of $\Phi$. 
We perform  a Fourier transformation on the $x$ coordinates, leaving the $\th$ coordinates,
\begin{equation}
\Phi(x)= \int \Phi_p e^{ip\cdot x} d^4p.
\end{equation} 
Then the superspace propagator can be derived from the quadratic part of the action,
\begin{eqnarray}
&&\langle 
\Phi_{p_{(1)}}{}^u{}_v \left(\th_{(1)}, \bar\th_{(1)}\right)
\Phi_{p_{(2)}}{}^r{}_s \left(\th_{(2)}, \bar\th_{(2)}\right)
\rangle
\nonumber\\
&=&
\frac{1}{(2\pi)^4}\delta^4(p_{(1)}+p_{(2)})\times
\delta^u{}_s \delta^r{}_v \times
\frac{i}{p_\m^2}
d^4_{(1)}
\delta^8(\th_{(1)}-\th_{(2)}).
\end{eqnarray}
Here $U(N)$ colour indices $u,v,\ldots$ are shown explicitly.

The cubic and quartic  vertices are read off 
from the interaction part of the action,
\begin{eqnarray}
iS_{\mbox{int}}& = &
\int d^4 x d^8 \th \Tr \Bigg(
-\frac{g}{6} 
\frac{\bar{d}^4}{\der_-^3} \Phi  \cdot
\bigg[ \Phi, 
\bar\der \Phi \bigg]_*
-\frac{g}{12} 
\frac{1}{\der_-} \Phi  \cdot
\bigg[ \frac{\bar{d}^4}{\der_-^2} \Phi, 
\der \frac{\bar{d}^4}{\der_-^2}\Phi \bigg]_*
\nonumber\\
&&+\frac{ig^2}{16} 
\frac{1}{\der_-} 
\bigg[ \Phi, 
\der_- \Phi \bigg]_*
\cdot 
\frac{1}{\der_-} 
\bigg[ \frac{\bar{d}^4}{\der_-^2}\Phi, 
\frac{\bar{d}^4}{\der_-} \Phi \bigg]_*
+\frac{ig^2}{32} 
\bigg[ \Phi, 
\frac{\bar{d}^4}{\der_-^2} \Phi \bigg]_*
\bigg[ \Phi, 
\frac{\bar{d}^4}{\der_-^2} \Phi \bigg]_*
\Bigg).
\end{eqnarray}
In momentum space this becomes 
\begin{eqnarray}
iS_{\mbox{int}}
&=&
\int d^8\th d^4k d^4p d^4q
(2\pi)^4 \delta^4(k+p+q)
e^{-\frac{i}{2}\left(k_\m C^{\m\n} l_\n + p_\m C^{\m\n} q_\n \right)  }
\nonumber\\
& &
\times
\Tr
\Bigg(
\frac{g}{6}
\frac{\bar{q}-\bar{p}}
{k_-^3}
\left(\bar{d}^4\Phi\right)_k
\Phi_p
\Phi_q
+
 \frac{g}{12}
\frac{p-q}{k_- p_-^2 q_-^2}
\Phi_k
\left(\bar{d}^4\Phi\right)_p
\left(\bar{d}^4\Phi\right)_q
\Bigg)
\nonumber\\
&+&
\int d^8\th d^4k d^4l d^4p d^4q
(2\pi)^4 \delta^4(k+l+p+q)
e^{-\frac{i}{2}\left(k_\m C^{\m\n} l_\n + p_\m C^{\m\n} q_\n \right)}
\nonumber\\
& &
\times
\Tr \Bigg(
 \frac{ig^2}{8} 
\frac{
k_- q_- +l_-p_-
}{
p_-^2 q_-^2 (p+q)_-^2
}
\Phi_k
\Phi_l
\left(\bar{d}^4\Phi\right)_p
\left(\bar{d}^4\Phi\right)_q
+
 \frac{ig^2}{16} 
\frac{1
}{
l_-^2 q_-^2 
}
\Phi_k
\left(\bar{d}^4\Phi\right)_l
\Phi_p
\left(\bar{d}^4\Phi\right)_q
\Bigg). 
\end{eqnarray}

We make it as a rule to write all vertices
in a way in which matrix products go in a counter clock-wise order.
In this convention, the diagrams can be written unambiguously
without the double-line notation, 
and we do not show the colour indices explicitly in the Feynman rules.  
One can easily recast 
the single-line diagrams in our convention
as diagrams written in the double-line convention,
which in turn is convenient to study 
the colour structure of the result.

As explained in the main text we perform the $\th$-integral 
by moving $d$'s and $\bar d$'s
via partial integrations,
using the identity \eqref{RFGSR} and
\begin{equation}
\{d^m, \bar{d}_n\} = \sqrt{2} i \delta^m{}_n \der_-.
\end{equation}
The Feynman rules are, 
\begin{eqnarray}
%propagator
\vcenter{\hbox{$\genfrac{}{}{0pt}{}{\includegraphics{vertex-0.mps}}{\rule{0pt}{1pt}}$}}
&=&
\frac{i}{p_\m^2}
\\
%3-Vertex with one $\bar{d}^4$'s
\vcenter{\hbox{\includegraphics{vertex-1.mps}}}
&=&
\frac{g}{6}
\frac{\bar{q}-\bar{p}}
{k_-^3}
e^{-\frac{i}{2} p_\m C^{\m\n} q_\n}
\\[5mm]
%3-Vertex with two $\bar{d}^4$'s
\vcenter{\hbox{\includegraphics{vertex-2.mps}}}
&=&\frac{g}{12}
\frac{p-q}{k_- p_-^2 q_-^2}
e^{-\frac{i}{2} p_\m C^{\m\n} q_\n}
\\[5mm]
%4-vertex with (no $\bar{d}^4$)-(no $\bar{d}^4$)-(with $\bar{d}^4$)-(with $\bar{d}^4$)
\vcenter{\hbox{\includegraphics{vertex-3.mps}}}
&=&
\frac{ig^2}{8} 
\frac{
k_- q_- +l_-p_-
}{
p_-^2 q_-^2 (p+q)_-^2
}
e^{-\frac{i}{2}\left(k_\m C^{\m\n} l_\n + p_\m C^{\m\n} q_\n \right)  }
\\[5mm]
%4-vertex with (no $\bar{d}^4$)-(with $\bar{d}^4$)-(no $\bar{d}^4$)-(with $\bar{d}^4$)
\vcenter{\hbox{\includegraphics{vertex-4.mps}}}
&=&
\frac{ig^2}{16}
\frac{1
}{
l_-^2 q_-^2 
}
e^{-\frac{i}{2}\left(k_\m C^{\m\n} l_\n + p_\m C^{\m\n} q_\n \right)  }
\end{eqnarray}
Here we used the convention
\begin{equation}
p=\frac{1}{\sqrt{2}}(p^1+ip^2), \quad \bar{p}=\frac{1}{\sqrt{2}}(p^1-ip^2).
\end{equation}
We use indices $\m, \n =0, 1, \ldots 3$ for four-vectors, 
and indices $i, j=1, 2$ for the transverse components.   
%%%%%%%%%%%%%%%%%%%%%%%%%%%%%%%%%%%%%%%%%
%%%%%%%%%%%%%%%%%%%%%%%%%%%%%%%%%%%%%%%%%
%%%%%%%%%%%%%%%%%%%%%%%%%%%%%%%%%%%%%%%%%
\subsection{One loop computation}
\label{RSSOneLoop}
%%%%%%%%%%%%%%%%%%%%%%%%%%%%%%%%%%%%%%%%%
%%%%%%%%%%%%%%%%%%%%%%%%%%%%%%%%%%%%%%%%%
%%%%%%%%%%%%%%%%%%%%%%%%%%%%%%%%%%%%%%%%%
We shall compute the two-point part of the 1PI effective action. 
Firstly we define the partition function with a source $J$ by 
\begin{equation}
Z[J]=\int e^{i S+ J\cdot \phi} \mathcal{D}\phi.
\end{equation}
Then by using the vacuum expectation value of $\phi$, 
we introduce the effective action $\Gamma$ as 
\begin{equation}
i \C= \log Z -J \cdot \langle \phi \rangle. 
\end{equation}
The effective action takes as its argument  
the vacuum expectation value of the quantum field.
For simplicity, we shall use in the following 
the same symbol $\Phi$ for the vacuum expectation
value of the superfield as the argument of the effective action.

\begin{figure}[htbp]
  \begin{center}
  \scalebox{0.3}{
   \includegraphics{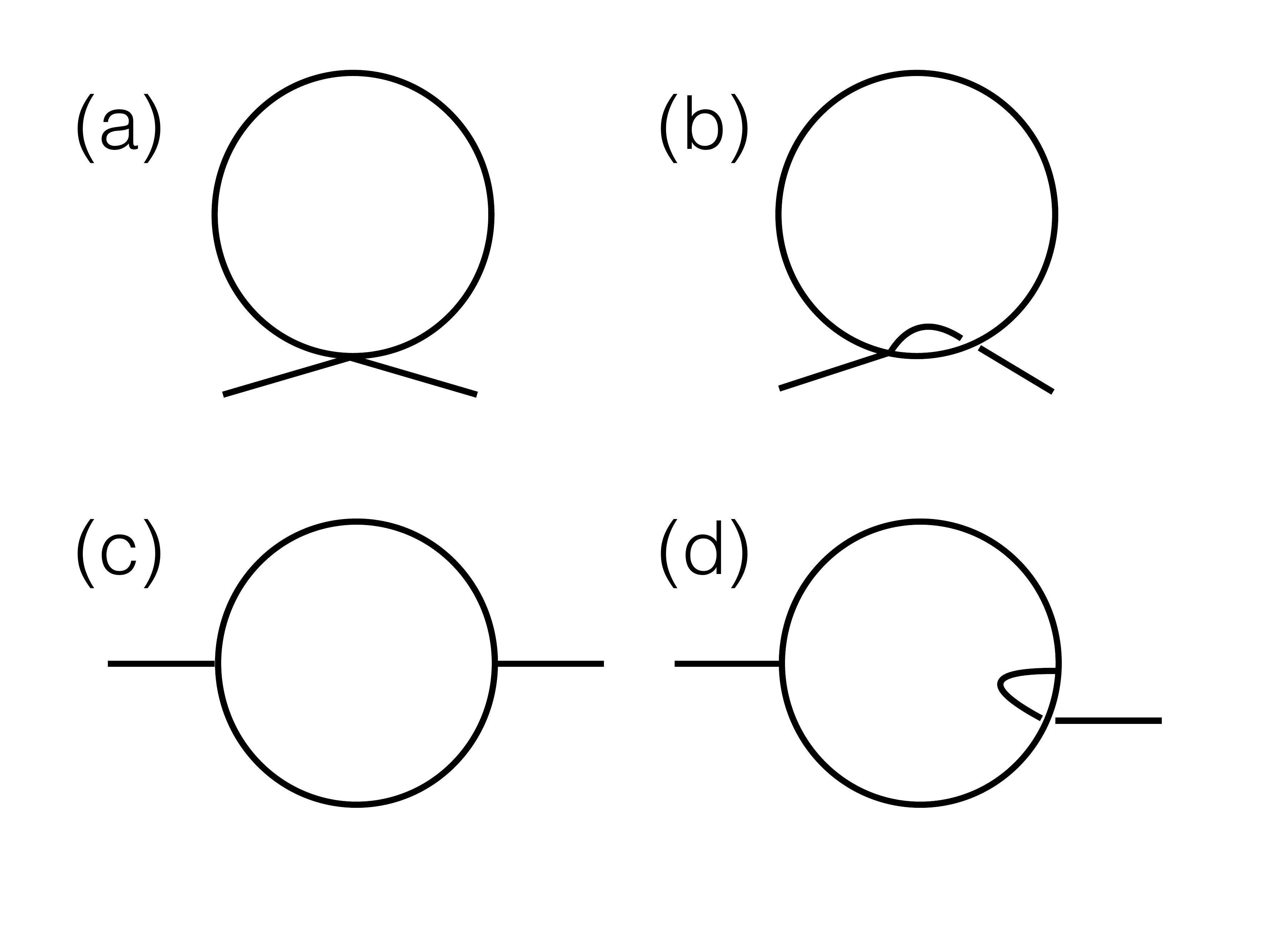}}
  \caption{One loop contributions to $i\Gamma$. 
%   (a) Omega diagram, planar; (b) Omega diagram, nonplanar;
%  (c) vacuum polarisation diagram, planar; (d) vacuum polarisation diagram, nonplanar. 
  }\label{fig:one-loop-diagrams}
  \end{center}
\end{figure}

One loop contributions to the two-point part of $i\Gamma$ come from 
the four diagrams in Fig.~\ref{fig:one-loop-diagrams}.

The contribution to $i \Gamma$ from the diagram 
(a) is 
\begin{equation}
\int d^4 k d^8 \th (2\pi)^4 \Tr (\bar{d}^4 \Phi)_k \Phi_{-k}
\times
\int\frac{d^4 p}{(2\pi)^4} g^2 N
\frac{1}{p_\m^2 (p_- +k_-)^2(p_--k_-)^2}
\end{equation}
which is derived by the superfield technique first introduced for 
$\mN=1$ superfields in \cite{RBGrisaruSiegelRocek}, 
see also \cite{RBThousandonelessons}.
For the case of the $\mN=4$ theory 
in the lightcone gauge, see \cite{RBBrinkLindgrenNilssonProof,
RBAKS1}.
The superficial degree of divergence is $-2$
in usual Lorentz invariant power counting. If only transverse momenta
are large, it is $0$. If only longitudinal momenta are large,
it is $-4$. 
We represent this by saying the superficial degree of divergence is
$(-2, 0, -4)$.
The contribution of this diagram is actually logarithmically divergent,
because of the region where transverse components are large and longitudinal 
components are finite.
Actually there are similar 
logarithmically divergent terms from (c), which cancel the divergent contribution.
In order to compute this integral, we rewrite the contribution as 
\begin{eqnarray}
&&\int d^4 k d^8 \th (2\pi)^4 Tr (\bar{d}^4 \Phi)_k \Phi_{-k}
\times
\int\frac{d^4 p}{(2\pi)^4} \left(\frac{g^2 N}{4}\right)
\times
\nonumber\\
&&
\left(
\frac{1}{k_-^2}\frac{1}{ p_\m^2  (p_-+k_-)^2}
+
\frac{1}{k_-^2}\frac{1}{ p_\m^2  (p_--k_-)^2}
+
\frac{1}{k_-^3}\frac{1}{ p_\m^2  (p_-+k_-)}
+
\frac{-1}{k_-^3}\frac{1}{ p_\m^2  (p_- - k_-)}
\right).
\end{eqnarray}
by using a partial fraction expansion.

The contribution from the diagram (b) is,
%Omega diagram, non-planar, fig. (b)
\begin{eqnarray}
&&\int d^4 k d^8 \th (2\pi)^4 \Tr (\bar{d}^4 \Phi)_k \Tr \Phi_{-k}
\times
\int\frac{d^4 p}{(2\pi)^4} \left(-g^2\right)
e^{i p_\m C^{\m\n} k_\n}
\frac{1}{p_\m^2 (p_- + k_-)^2 (p_- - k_-)^2}
\nonumber\\
&=&\int d^4 k d^8 \th (2\pi)^4 \Tr (\bar{d}^4 \Phi)_k \Tr \Phi_{-k}
\times
\int\frac{d^4 p}{(2\pi)^4} \left(\frac{-g^2}{4}\right)
e^{i p_\m C^{\m\n} k_\n}
\times
\nonumber\\
&&
\left(
\frac{1}{ k_-^2}
\frac{1}{p_\m^2 (p_- + k_-)^2}
+
\frac{1}{ k_-^2 }
\frac{1}{p_\m^2(p_- - k_-)^2}
+
\frac{1}{ k_-^3}
\frac{1}{p_\m^2 (p_- + k_-)}
+
\frac{-1}{ k_-^3}
\frac{1}{p_\m^2 (p_-  - k_-)}
\right)
\end{eqnarray}
We note that the contribution from the planar diagram (a) 
and the non-planar diagram (b) differs  simply by a  factor $-\frac{1}{N}$  
and the phase factor $e^{i p_\m C^{\m\n} k_\n}$ at the level of the integrand, as it should be.
The superficial degree of divergence is the same as the 
corresponding planar diagram (a), i.e. $(-2,0,-4)$.

The contribution from the diagram (c) is 
\begin{eqnarray}
&&\int d^4 k d^8 \th (2\pi)^4 \Tr (\bar{d}^4 \Phi)_k \Phi_{-k}
\times
\int\frac{d^4 p}{(2\pi)^4} \left(-\frac{g^2 N}{18}\right)
\frac{1}{p_\m^2 (p+k)_\m^2 k_-^2 p_-^2(p_-+k_-)^2}
\nonumber\\
&&\qquad\quad \times
|(2p+k)k_- +(p-k)(p+k)_- - (p+2k)p_-|^2
\nonumber\\
&=&
\int d^4 k d^8 \th (2\pi)^4 \Tr (\bar{d}^4 \Phi)_k \Phi_{-k}
\times
\int\frac{d^4 p}{(2\pi)^4} \left(-\frac{g^2 N}{4}\right)
\frac{
p_i^2 k_-^2 + k_i^2 p_-^2 -2k_i p_i k_- p_-
}{p_\m^2 (p+k)_\m^2 k_-^2 p_-^2(p_-+k_-)^2
}
\end{eqnarray}
The superficial degree of divergence is again $(-2,0,-4)$.
This can be rewritten as,
\begin{eqnarray}
&&\int d^4 k d^8 \th (2\pi)^4 \Tr (\bar{d}^4 \Phi)_k \Phi_{-k}
\times
\int\frac{d^4 p}{(2\pi)^4} \left(-\frac{g^2 N}{4}\right)
\times
\nonumber\\
&&
\bigg(
\frac{1}{k_-^2} 
\frac{1}{(p+k)_\m^2 p_-^2}
+
\frac{1}{k_-^2}
\frac{1}{p_\m^2(p_-+k_-)^2}
\nonumber\\
&&
+
\frac{-1}{k_-^3} 
\frac{1}{(p+k)_\m^2 p_-}
+
\frac{1}{k_-^3}
\frac{1}{p_\m^2 (p_-+k_-)}
+
\frac{1}{k_-^3} 
\frac{1}{(p+k)_\m^2 (p_-+k_-)}
+
\frac{-1}{k_-^3}
\frac{1}{p_\m^2 p_-}
\nonumber\\
&&
+
\frac{k_\m^2}{k_-^3} 
\frac{1}{p_\m^2(p+k)_\m^2 p_-}
+
\frac{-k_\m^2}{k_-^3} \frac{1}{p_\m^2(p+k)_\m^2 (p_-+k_-)}
\bigg).
\end{eqnarray}
Here the key steps in the manipulation are 
\begin{eqnarray}
&&
\frac{
p_i^2 k_-^2 + k_i^2 p_-^2 -2k_i p_i k_- p_-
}{p_\m^2 (p+k)_\m^2 k_-^2 p_-^2(p_-+k_-)^2}
=
\frac{
k_-(p_-+k_-) p_i^2 
- k_-p_- (p+k)_i^2
+p_-(p_-+k_-) k_i^2
}
{p_\m^2 (p+k)_\m^2 k_-^2 p_-^2 (p_-+k_-)^2}
\nonumber\\
&=&
\frac{1}{
k_-}
\frac{
 p_i^2 
}
{p_\m^2 (p+k)_\m^2 p_-^2 (p_-+k_-)}
+
\frac{1}{
k_-}
\frac{
- (p+k)_i^2
}
{p_\m^2 (p+k)_\m^2 p_- (p_-+k_-)^2}
+
\frac{1}{
k_-^2}
\frac{
 k_i^2
}
{p_\m^2 (p+k)_\m^2  p_- (p_-+k_-)}
\nonumber\\
&=&
\frac{1}{
k_-}
\frac{
p_\mu^2+2p_+p_-
}
{p_\m^2 (p+k)_\m^2 p_-^2 (p_-+k_-)}
+
\frac{1}{
k_-}
\frac{
- (p+k)_\m^2 -2(p+k)_+(p+k)_-
}
{p_\m^2 (p+k)_\m^2 p_- (p_-+k_-)^2}
+
\frac{1}{
k_-^2}
\frac{
 k_i^2
}
{p_\m^2 (p+k)_\m^2  p_- (p_-+k_-)}
\nonumber\\
&=&
\frac{1}{
k_-}
\frac{
1
}
{ (p+k)_\m^2 p_-^2 (p_-+k_-)}
+
\frac{-1}{
k_-}
\frac{
1
}
{p_\m^2  p_- (p_-+k_-)^2}
\nonumber\\
&&
+
\frac{1}{
k_-}
\frac{
2p_+
}
{p_\m^2 (p+k)_\m^2 p_- (p_-+k_-)}
+
\frac{-1}{
k_-}
\frac{
2(p+k)_+
}
{p_\m^2 (p+k)_\m^2 p_- (p_-+k_-)}
+
\frac{1}{
k_-^2}
\frac{
 k_i^2
}
{p_\m^2 (p+k)_\m^2  p_- (p_-+k_-)}
\nonumber\\
%END
%BEGIN
&=&
\frac{1}{
k_-}
\frac{
1
}
{ (p+k)_\m^2 p_-^2 (p_-+k_-)}
+
\frac{-1}{
k_-}
\frac{
1
}
{p_\m^2  p_- (p_-+k_-)^2}
+
\frac{k_\m^2}{
k_-^2}
\frac{
1
}
{p_\m^2 (p+k)_\m^2  p_- (p_-+k_-)}.
\end{eqnarray}
%END
The last equality involves 
cancellations between terms proportional to $p_+$. %Looks very much like a face

The contribution from 
the diagram (d) is
%Vac.-pol.-diagram, non-planar, fig. (d) 
\begin{equation}
\int d^4 k d^8 \th (2\pi)^4 \Tr (\bar{d}^4 \Phi)_k \Tr \Phi_{-k}
\times
\int\frac{d^4 p}{(2\pi)^4} \left(\frac{g^2 }{4}\right)
e^{i p_\m C^{\m\n} k_\n}
\frac{
p_i^2 k_-^2 + k_i^2 p_-^2 -2k_i p_i k_- p_-
}{p_\m^2 (p+k)_\m^2 k_-^2 p_-^2(p_-+k_-)^2},
\end{equation}
whose superficial degree of divergence is $(-2,0,-4)$.
Again, the contribution of (c) and (d) is related simply by a factor of $-1/N$ and the non-commutative phase 
at the level of the integrand.
By a computation  similar to the case of (c), 
this reduces to
\begin{eqnarray}
&&\int d^4 k d^8 \th (2\pi)^4 \Tr (\bar{d}^4 \Phi)_k \Tr \Phi_{-k}
\times
\int\frac{d^4 p}{(2\pi)^4} \left(\frac{g^2 }{4}\right)
e^{i p_\m C^{\m\n} k_\n}
\times
\nonumber\\
&&
\bigg(
\frac{1}{k_-^2} 
\frac{1}{(p+k)_\m^2 p_-^2}
+
\frac{1}{k_-^2}
\frac{1}{p_\m^2(p_-+k_-)^2}
\nonumber\\
&&
+
\frac{-1}{k_-^3} 
\frac{1}{(p+k)_\m^2 p_-}
+
\frac{1}{k_-^3}
\frac{1}{p_\m^2 (p_-+k_-)}
+
\frac{1}{k_-^3} 
\frac{1}{(p+k)_\m^2 (p_-+k_-)}
+
\frac{-1}{k_-^3}
\frac{1}{p_\m^2 p_-}
\nonumber\\
&&
+
\frac{k_\m^2}{k_-^3} 
\frac{1}{p_\m^2(p+k)_\m^2 p_-}
+
\frac{-k_\m^2}{k_-^3} \frac{1}{p_\m^2(p+k)_\m^2 (p_-+k_-)}
\bigg).
\end{eqnarray}

Combining the planar contributions
from (a) and (c), we obtain
\begin{eqnarray}
&&\int d^4 k d^8 \th (2\pi)^4 \Tr (\bar{d}^4 \Phi)_k \Phi_{-k}
\times
\int\frac{d^4 p}{(2\pi)^4} \left(-\frac{g^2 N}{4}\right)
\times
\nonumber\\
&&
\bigg(
\frac{-1}{ k_-^2 }
\frac{1}{p_\m^2(p_- - k_-)^2}
+
\frac{1}{k_-^2} 
\frac{1}{(p+k)_\m^2 p_-^2}
\nonumber\\
&&
+
\frac{1}{ k_-^3}
\frac{1}{p_\m^2 (p_-  - k_-)}
+
\frac{-1}{k_-^3} 
\frac{1}{(p+k)_\m^2 p_-}
+
\frac{1}{k_-^3} 
\frac{1}{(p+k)_\m^2 (p_-+k_-)}
+
\frac{-1}{k_-^3}
\frac{1}{p_\m^2 p_-}
\nonumber\\
&&
+
\frac{k_\m^2}{k_-^3} 
\frac{1}{p_\m^2(p+k)_\m^2 p_-}
+
\frac{-k_\m^2}{k_-^3} \frac{1}{p_\m^2(p+k)_\m^2 (p_-+k_-)}
\bigg).
\end{eqnarray}
By shifting integration variables appropriately,~\footnote{
Power-counting shows that the divergences of individual terms are at most linear. 
Furthermore, 
the linearly divergent parts vanishes because of rotational symmetry.
Thus all the terms are at most logarithmically divergent, and the
shifting of integration variables should be legitimate.
}
one can show 
that only the last two terms
in the parentheses remain. Hence we have,  
\begin{eqnarray}
&&\int d^4 k d^8 \th (2\pi)^4 \Tr (\bar{d}^4 \Phi)_k \Phi_{-k}
\times
\int\frac{d^4 p}{(2\pi)^4} \left(-\frac{g^2 N}{4}\right)
\times
\nonumber\\
&&
\left(
\frac{k_\m^2}{k_-^3} 
\frac{1}{p_\m^2(p+k)_\m^2 p_-}
+
\frac{-k_\m^2}{k_-^3} \frac{1}{p_\m^2(p+k)_\m^2 (p_-+k_-)}
\right).
\end{eqnarray}
The superficial degree of divergence is $(-1,-2,-3)$.
Thus we see that the integral is finite due to cancellations
between diagrams (a) and (c).
The integral can be computed by the method explained in the next subsection.
Using (\ref{RFIPlanar}),
this can be rewritten as, 
\begin{equation}
\int d^4 k d^8 \th (2\pi)^4 \Tr (\bar{d}^4 \Phi)_k \Phi_{-k}
\times
\left(-\frac{g^2 N}{2}\right)
\times 
 i \pi^{2} \frac{1}{k_-} 
\int_0^1 dt
\frac{1}{t}
\log{
\frac{
 t(1-t) k_\mu^2 
}{
(t k_\mu^2 -t^2 k_i^2) 
}
}.
\label{RFPlanarTwoPointFinal}
\end{equation}

For the combined non-planar contribution to $i \Gamma$ from (b) and (d),  
we can perform the same shift without changing the phase factor $e^{i p_\m C^{\m\n} k_\n}$. 
Then we obtain 
\begin{eqnarray}
&&\int d^4 k d^8 \th (2\pi)^4 \Tr (\bar{d}^4 \Phi)_k \Tr \Phi_{-k}
\times
\int\frac{d^4 p}{(2\pi)^4} \left(\frac{g^2}{4}\right)
e^{i p_\m C^{\m\n} k_\n}
\times
\nonumber\\
&&
\bigg(
\frac{k_\m^2}{k_-^3} 
\frac{1}{p_\m^2(p+k)_\m^2 p_-}
+
\frac{-k_\m^2}{k_-^3} \frac{1}{p_\m^2(p+k)_\m^2 (p_-+k_-)}
\bigg). 
\label{(b)and(d)}
\end{eqnarray}
The superficial degree of divergence is $(-1,-2,-3)$, 
and the integral is finite by cancellation between diagrams (b) and (d). 
Therefore, the commutative limit should be continuous. 
In order to see it explicitly, we rewrite this expression by using the Bessel function of the second kind, 
\begin{eqnarray}
\eqref{(b)and(d)}
&=& 
\int d^4 k d^8 \th (2\pi)^4 \Tr (\bar{d}^4 \Phi)_k \Tr \Phi_{-k}
\times
\left(\frac{g^2}{2}\right)
\times 
\nonumber\\
&&
\left(
- 2 i \pi^{2} \frac{1}{k_-} 
\int_0^1 dt
\frac{1}{t}
\left(
K_0\left(
\sqrt{
t(1-t)k_\m^2 \tk_i^2
}
\right)
-
K_0\left(
\sqrt{
(t k_\m^2 - t^2 k_i^2) \tk_i^2
}
\right)
\right)
\right), 
\end{eqnarray}
by using \eqref{RFI}. 
Here $\tilde{k}^\mu=C^{\mu\nu}k_\nu$. 
If one takes $C\rightarrow 0$ using 
the behaviour of 
$K_0(x)$ around $x\sim 0$,
\eqref{RFBesselAsymptotics},
this becomes
\begin{equation}
=\int d^4 k d^8 \th (2\pi)^4 \Tr (\bar{d}^4 \Phi)_k \Tr \Phi_{-k}
\times
\left(\frac{g^2}{2}\right)
\times 
\end{equation}
\begin{equation}
\left(
i \pi^{2} \frac{1}{k_-} 
\int_0^1 dt
\frac{1}{t}
\log{ \frac{
(1-t)k_\m^2 
}
{
(k_\m^2 - t k_i^2) 
}
}
\right). 
\end{equation}
This is $-\frac{1}{N}$ times the planar contribution (\ref{RFPlanarTwoPointFinal}).
Thus we have confirmed the continuity explicitly. 

%%%%%%%%%%%%%%%%%%%%%%%%%%%%%%%%%%%%%%
%%%%%%%%%%%%%%%%%%%%%%%%%%%%%%%%%%%%%%
%%%%%%%%%%%%%%%%%%%%%%%%%%%%%%%%%%%%%%
\subsection{Some useful formulae}
\label{RSSNiceFormulae}
%%%%%%%%%%%%%%%%%%%%%%%%%%%%%%%%%%%%%%
%%%%%%%%%%%%%%%%%%%%%%%%%%%%%%%%%%%%%%
%%%%%%%%%%%%%%%%%%%%%%%%%%%%%%%%%%%%%%
\subsubsection{Feynman integrals in lightcone gauge}
%%%%%%%%%%%%%%%%%%%%%%%%%%%%%%%%%%%%%%
%%%%%%%%%%%%%%%%%%%%%%%%%%%%%%%%%%%%%%
%%%%%%%%%%%%%%%%%%%%%%%%%%%%%%%%%%%%%%
In order to evaluate Feynman integrals in the lightcone gauge,
we follow the method explained in 
\cite{RBCapperDulwichLitvak}.
We start from the following integral \cite{RBCapperDulwichLitvak},
\begin{equation}
\int dp_+ dp_- d^{2-\e} p
\frac{1}{p_-} 
e^{-i \a p_\m p^\m -2 i \r_\m p^\m}
=
- \frac{1}{\r_-} \pi^{2-\frac{\e}{2}} (i\a)^{\frac{\e}{2}-1}
\left(
e^{i \frac{1}{\a} \r_\m \r^\m}
-
e^{i \frac{1}{\a} \r_i \r^i}
\right).
\label{RFIntegralGeneratingPowerOne}
\end{equation}
The above formula can be derived by performing the 
Wick rotation (which is allowed due to our use of the Mandelstam prescription
for factors such as $\frac{1}{p_-}$), 
and evaluating the integral 
over the longitudinal space and the transverse space successively. 
In this expression, the dimensional regularisation is used for the transverse dimensions
(i.e. the dimension of the transverse space 
is formally altered from $2$ to $2-\e$). 
For our purpose it is not necessary, and hence we will set $\e=0$ from now on. 

The details regarding the Wick rotation are as follows.
The integration contour for $p_0$ on the real axis from left to right is
rotated counter-clockwise by $\pi/2$, 
such that it coincides with the imaginary axis. 
It is convenient to define $i p_0^E=p_0$. 
After the rotation, $p_0^E$ is real, and $\int dp_0$ is
replaced by $i \int dp^E_0$. 
The parameter 
$\rho_\m$ can be either real, pure imaginary, complex
for all components. 
The integral is defined if $\mbox{Im} \alpha <0$.
The integration contour over the parameter $\alpha$
is rotated clockwise on the complex plane by $\pi/2$ later.

We define the integral $I$ 
\begin{equation}
I=\int d^{4} p 
\frac{1}{p_\m^2(p+k)_\m^2} \frac{1}{p_-}
e^{i p_\m C^{\m\n} k_\n},
\end{equation}
and also consider $I_{\mbox{planar}}$  defined by
\begin{equation}
I_{\mbox{planar}}=\int d^{4} p 
\frac{1}{p_\m^2(p+k)_\m^2} \frac{1}{p_-}.
\end{equation}
Both of them are convergent.

By using a Feynman parameter $t$, $I$ can be rewritten as
\begin{equation}
I=\int d^{4} p 
\int_0^1 dt
\frac{1}
{
\left(
(p+tk)_\mu^2
+
t(1-t) k_\mu^2
\right)^2
}
\frac{1}{p_-}
e^{i p_\m C^{\m\n} k_\n}
\end{equation}
By using a Schwinger parameter $\a$, this becomes,
\begin{eqnarray}
I&=&\int d^{4} p 
\int_0^1 dt
(-1) \int_0^{+\infty} d\a \a
e^{
-i \a \left(
(p+tk)_\mu^2
+
t(1-t) k_\mu^2
\right)
}
\frac{1}{p_-}
e^{i p_\m C^{\m\n} k_\n}
\nonumber\\
&=&
(-1)
\int d^{4} p 
\int_0^1 dt
\int_0^{+\infty} d\a \a
e^{
-i\a \left(
p_\m^2
+2tp_\m k^\m
+
t k_\mu^2
\right)
}
\frac{1}{p_-}
e^{i p_\m C^{\m\n} k_\n}
\nonumber\\
&=&
(-1)\int d^{4} p 
\int_0^1 dt
\int_0^{+\infty} d\a \a
e^{
-i \a 
p_\m^2
}
e^{ -2i p_\m
\left(
\a t k^\m
-\frac{1}{2} C^{\m\n} k_\n
\right)
}
e^{
-i \a t k_\mu^2
}
\frac{1}{p_-}.
\end{eqnarray}
We now apply \eqref{RFIntegralGeneratingPowerOne} with
\begin{equation}
\rho^\m=
\a t k^\m
-\frac{1}{2} C^{\m\n} k_\n
=
\a t k^\m
-\hk^\m.
\end{equation}
We use the notation
\begin{equation}
\tk^\m=2\hk^\m= C^{\m\n} k_\n
\end{equation}
for brevity.
Then, 
\begin{equation}
I=
(-1)
\int_0^1 dt
\int_0^{+\infty} d\a \a
\left(
- \frac{1}{\r_-} \pi^{2} (i\a)^{-1}
\left(
e^{i \frac{1}{\a} \r_\m \r^\m}
-
e^{i \frac{1}{\a} \r_i \r^i}
\right)
e^{
-i \a t k_\mu^2
}
\right).
\end{equation}
Since the only non-zero components of $C$ are $C^{ij}$, we have
\begin{equation}
\rho_\mu^2
=
\a^2 t^2 k_\m^2
+\hk_i^2, \quad
\rho_i^2
=
\a^2 t^2 k_i^2
+\hk_i^2,\quad
\rho_-=\a t k_-.
\end{equation}
We note that $k_i \hk_i=0$.
Hence, we obtain
\begin{equation}
I=
 \pi^{2} 
\int_0^1 dt
\int_0^{+\infty} d\a \a
\frac{1}{\a t k_-}
(i\a)^{-1}
\left(
e^{
-i \a t(1-t) k_\mu^2 +i \frac{1}{\a} \hk_i^2
}
-
e^{
-i \a (t k_\mu^2 -t^2 k_i^2) +i \frac{1}{\a} \hk_i^2
}
\right).
\end{equation}
We now rotate the integration contour of $\a$ by writing,
$u=i\a$.~\footnote{
We assume that 
the external momenta $k_\m$ 
are analytically continued appropriately 
in order to avoid any problem which may occur in the Wick rotation of the contour.}
\begin{eqnarray}
I&=&
- \pi^{2} 
\int_0^1 dt
\int_0^{+\infty} du u
u^{-1}
\frac{1}{-i t u k_-}
\left(
e^{
-u t(1-t) k_\mu^2 - \frac{1}{u} \hk_i^2
}
-
e^{
-u (t k_\mu^2 -t^2 k_i^2) - \frac{1}{u} \hk_i^2
}
\right)
\nonumber\\
&=&
- i \pi^{2} \frac{1}{k_-} 
\int_0^1 dt
\frac{1}{t}
\int_0^{+\infty} du 
\frac{1}{u}
\left(
e^{
-u t(1-t) k_\mu^2 - \frac{1}{4u} \tk_i^2
}
-
e^{
-u (t k_\mu^2 -t^2 k_i^2) - \frac{1}{4u} \tk_i^2
}
\right).
\end{eqnarray}
Applying the integral representation of the Bessel function, \eqref{RFBesselIntegral}, 
we finally obtain 
\begin{equation}
I=
- 2 i \pi^{2} \frac{1}{k_-} 
\int_0^1 dt
\frac{1}{t}
\left(
K_0\left(
\sqrt{
t(1-t)k_\m^2 \tk_i^2
}
\right)
-
K_0\left(
\sqrt{
(t k_\m^2 - t^2 k_i^2) \tk_i^2
}
\right)
\right).
\label{RFI}
\end{equation}
This is an odd function with respect to $k_\m$.

We proceed similarly for the $I_{\mbox{planar}}$,
\begin{equation}
I_{\mbox{planar}}=
- i \pi^{2} \frac{1}{k_-} 
\int_0^1 dt
\frac{1}{t}
\int_0^{+\infty} du 
\frac{1}{u}
\left(
e^{
-u t(1-t) k_\mu^2 
}
-
e^{
-u (t k_\mu^2 -t^2 k_i^2) 
}
\right).
\label{RFIPlanar}
\end{equation}
In order to evaluate this, we consider 
\begin{equation}
\int_0^{+\infty} du 
\frac{1}{u}
\left(
e^{
-A u  
}
-
e^{
-B u  
}
\right).
\end{equation}
This integral is convergent as the two terms 
cancel each other when $u\sim 0$.
By partial  integration, we have
\begin{eqnarray}
\int_0^{+\infty} du 
\frac{1}{u}
\left(
e^{
-A u  
}
-
e^{
-B u  
}
\right)
&=&
\left.
\log u \left(e^{
-A u  
}
-
e^{
-B u  }\right) \right |^{+\infty}_0
-
\int_0^{+\infty} du 
\log u
\left(
-A
e^{ -A u  }
- (-B)
e^{ -B u  }
\right)
\nonumber\\
&=&-\log{\frac{A}{B}}.
\end{eqnarray}
In order to show the last equality,
notice that the two integrals in the second expression are separately convergent,
and perform the change of variables $v=Au$ and $v=Bu$ for them, respectively.

Finally, we have
\begin{equation}
I_{\mbox{planar}}=
 i \pi^{2} \frac{1}{k_-} 
\int_0^1 dt
\frac{1}{t}
\log{
\frac{
 (1-t) k_\mu^2 
}{
(k_\mu^2 -t k_i^2) 
}
}.
\end{equation}
The argument of the $\log$ function is $1$ for $t\sim 0$,
so that the integral over the Feynman parameter is convergent.
This is an odd function with respect to $k_\m$.

\subsubsection{Bessel functions}
The modified Bessel functions of the second kind $K_\nu(x)$
have an integral representation
\footnote{
We follow the convention of \cite{RBDLMF}.}
\begin{equation}
K_\n(\sqrt{\b\c})
=
2^{\n-1}
\sqrt{\frac{\c}{\b}}^\n
\int_0^{+\infty}
e^{-\frac{\b}{4u}-\c u}
u^{\n -1}
du.
\label{RFBesselIntegral}
\end{equation}
Here we assume $\b>0, \c>0$.

Their behaviours at $x\sim 0$ are governed, for integer valued $\n$,  by
the expansion
\begin{eqnarray}
K_{n}(x)
&=&
\frac{1}{2}\left(\frac{x}{2}\right)^{-n}
\sum_{k=0}^{n-1}\frac{(n-k-1)!}{k!}\left(-\frac{x^2}{4}\right)^{k}
+
(-1)^{n+1}\log\left(\frac{x}{2}\right)
I_{n}\left(x\right)
\nonumber\\
&+&
(-1)^{n}\frac{1}{2}\left(\frac{x}{2}\right)^{n}
\sum_{k=0}^{\infty}
\frac{1}{(n+k)!k!}
\left(
\psi\left(k+1\right)+\psi(n+k+1)\right)
\left(\frac{x^2}{4}\right)^{k},
\end{eqnarray}
where $I_\n(x)$ is defined by
\begin{equation}
I_\n(x)=\left(\frac{x}{2}\right)^\n
\sum_{n=0}^{+\infty} 
\frac{\left(\frac{x^2}{4}\right)^n} {n! \Gamma(n+\n+1)}
\end{equation}
and $\psi(x)=\frac{\Gamma'(x)}{\Gamma(x)}$. Specifically, we have
\begin{eqnarray}
K_0(x)&=& -\log{\frac{x}{2}} \times I_0(x)
+\sum_{k=0}^{+\infty} \frac{1}{(k!)^2} \psi(k+1 )
\left(\frac{x^2}{4}\right)^k,
\\
I_0(x)&=&
\sum_{k=0}^{+\infty} \frac{1}{(k!)^2} 
\left(\frac{x^2}{4}\right)^k.
\end{eqnarray}
Writing down the first few terms, we have
\begin{equation}
K_0(x)
=
-\log{\frac{x}{2}} \times \left(1+\frac{x^2}{4}+\cdots\right)
+
\left(\psi(1)+\psi(2)\frac{x^2}{4}+\cdots\right).
\label{RFBesselAsymptotics}
\end{equation}

%%%%%%%%%%%%%%%%%%%%%%%%%%%%%%%%%%%%%%%%%%%%%%%%%%%%%%%%%
%%%%%%%%%%%%%%%%%%%%%%%%%%%%%%%%%%%%%%%%%%%%%%%%%%%%%%%%%

\end{document}